\documentclass[preprint,12pt]{elsarticle}

\usepackage{subfigure}
\usepackage{amsmath}
\usepackage[ruled,linesnumbered,slide]{algorithm2e}
\usepackage{slashbox}
\usepackage{url}
\usepackage{caption}
\usepackage{rotating}
\usepackage{ragged2e}
\usepackage{soul}
\usepackage{fnpos}

\usepackage{amssymb}

\journal{Journal of Network and Computer Applications}

\begin{document}

\begin{frontmatter}



\title{A Reliable and Cost-Efficient Auto-Scaling System for Web Applications Using Heterogeneous Spot Instances}

\author{Chenhao Qu, Rodrigo N. Calheiros, and Rajkumar Buyya}

\address{Cloud Computing and Distributed Systems (CLOUDS) Laboratory}
\address{Department of Computing and Information Systems}
\address{The University of Melbourne, Australia}

\begin{abstract}
Cloud providers sell their idle capacity on markets through an auction-like mechanism to increase their return on investment. The instances sold in this way are called spot instances. In spite that spot instances are usually 90\% cheaper than on-demand instances, they can be terminated by provider when their bidding prices are lower than market prices. Thus, they are largely used to provision fault-tolerant applications only. In this paper, we explore how to utilize spot instances to provision web applications, which are usually considered availability-critical. The idea is to take advantage of differences in price among various types of spot instances to reach both high availability and significant cost saving. We first propose a fault-tolerant model for web applications provisioned by spot instances. Based on that, we devise novel cost-efficient auto-scaling polices that comply with the defined fault-tolerant semantics for hourly billed cloud markets. We implemented the proposed model and policies both on a simulation testbed for repeatable validation and Amazon EC2. The experiments on the simulation testbed and EC2 show that the proposed approach can greatly reduce resource cost and still achieve satisfactory Quality of Service (QoS) in terms of response time and availability.
\end{abstract}


\end{frontmatter}


\section{Introduction}\label{sec:introduction}

There are three common pricing models in current Infrastructure-as-a-service (IaaS) cloud providers, namely \emph{on-demand}, in which acquired virtual machines (VMs) are charged periodically with fixed rates, \emph{reservation}, where users pay an amount of up-front fee for each VM to secure availability of usage and cheaper price within a certain contract period, and the \emph{spot}.

The spot pricing model was introduced by Amazon to sell their spare capacity in open market through an auction-like mechanism. The provider dynamically sets the market price of each VM type according to real-time demand and supply. To participate in the market, a cloud user needs to give a bid specifying number of instances for the type of VM he wants to acquire and the maximum unit price he is willing to pay. If the bidding price exceeds the current market price, the bid is fulfilled. After getting the required spot VMs, the user only pays the current market prices no matter how much he actually bids, which results in significant cost saving compared to VMs billed in on-demand prices (usually only 10\% to 20\% of the latter) \cite{Amazon}. However, obtained spot VMs will be terminated by cloud provider whenever their market prices rise beyond the bidding prices.

Such model is ideal for fault-tolerant and non-time-critical applications such as scientific computing, big data analytics, and media processing applications. On the other hand, it is generally believed that availability- and time-critical applications, like web applications, are not suitable to be deployed on spot instances.
 
Adversely in this paper, we illustrate that, with effective fault-tolerant mechanism and carefully designed policies that comply with the fault-tolerant semantics, it is also possible to reliably scale web applications using spot instances to reach both high QoS and significant cost saving.

\begin{figure}
\centering
\includegraphics[width=5in]{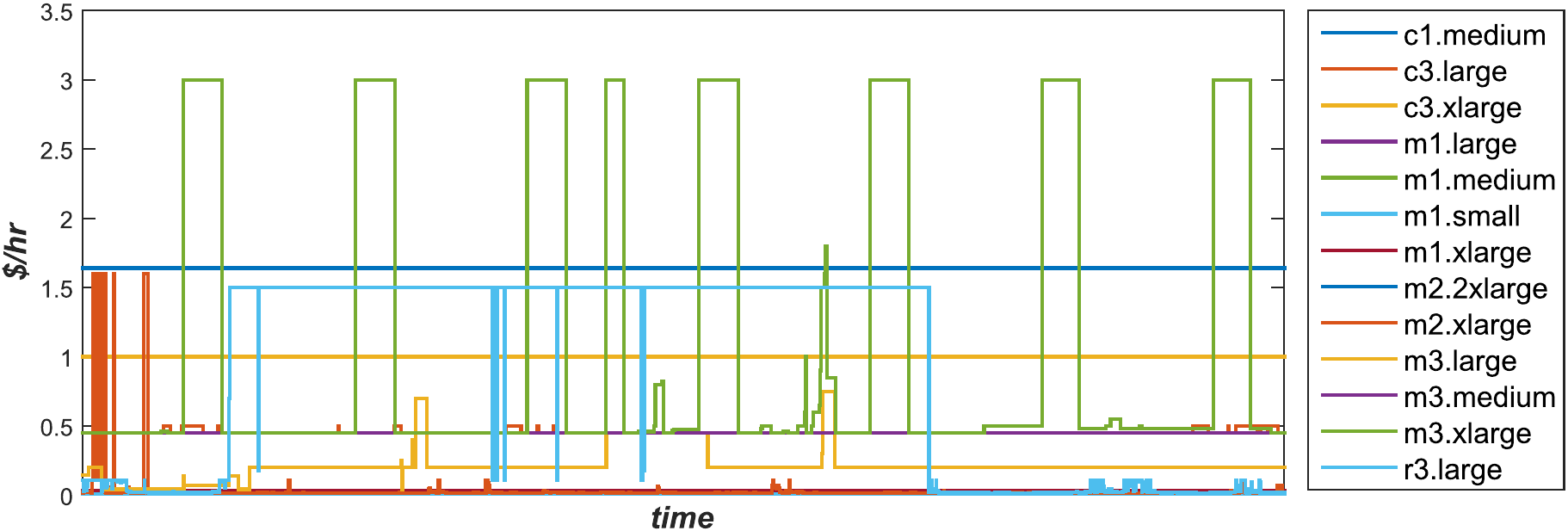}
\caption{One week spot price history from March 2nd 2015 18:00:00 GMT in Amazon EC2's \emph{\textbf{us-east-1d}} Availability Zone}
\label{fig:spot_price_history}
\end{figure}

Spot market is similar to a stock market that, though possibly following the general trends, each listed item has its distinctive market behaviour according to its own supply and demand. In this kind of market, often price differences appear with some types of instances sold in expensive prices due to high demand, while some remaining unfavoured leading to attractive deals. Figure \ref{fig:spot_price_history} depicts a period of Amazon EC2's spot market history. Within this time frame, there were always some spot types sold in discounted prices. By exploiting the diversity in this market, cloud users can utilize spot instances as long as possible to further reduce their cost. Recently, Amazon introduced the Spot Fleet API \cite{Amazona}, which allows users to bid for a pool of resources at once. The provision of resources is automatically managed by Amazon using combination of spot instances with lowest cost. However, it still lacks fault-tolerant capability to avoid availability and performance impact caused by sudden termination of spot instances, and thus, is not suitable to provision web applications.

To fill in this gap, we aim to build a solution to cater this need. We proposed a reliable auto-scaling system for web applications using heterogeneous spot instances along with on-demand instances. Our approach not only greatly reduces financial cost of using cloud resources, but also ensures high availability and low response time, even when some types of spot VMs are terminated unexpectedly by cloud provider simultaneously or consecutively within a short period of time.

The \textbf{main contributions} of this paper are: 

\begin{itemize}
\item a fault-tolerant model for web applications provisioned by spot instances;
\item cost-efficient auto-scaling policies that comply with the defined fault-tolerant semantics using heterogeneous spot instances;
\item event-driven prototype implementations of the proposed auto-scaling system on CloudSim \cite{Calheiros2011} and Amazon EC2 platform;
\item performance evaluations through both repeatable simulation studies based on historical data and real experiments on Amazon EC2;
\end{itemize}

The remainder of the paper is organized as follows. We first model our problem in Section \ref{sec:system_model}. In section \ref{sec:scaling_policies}, we propose the base auto-scaling policies using heterogeneous spot instances under hourly billed context. Section \ref{sec:optimizations} explains the optimizations we proposed on the initial polices. Section \ref{sec:implementation} briefly introduces our prototype implementations. We present and analyze the results of the performance evaluations in Section \ref{sec:performance_evalutaion} and discuss the related works in Section \ref{sec:related_work}. Finally, we conclude the paper and vision our future work.

\begin{figure}
\centering
\includegraphics[width=4in]{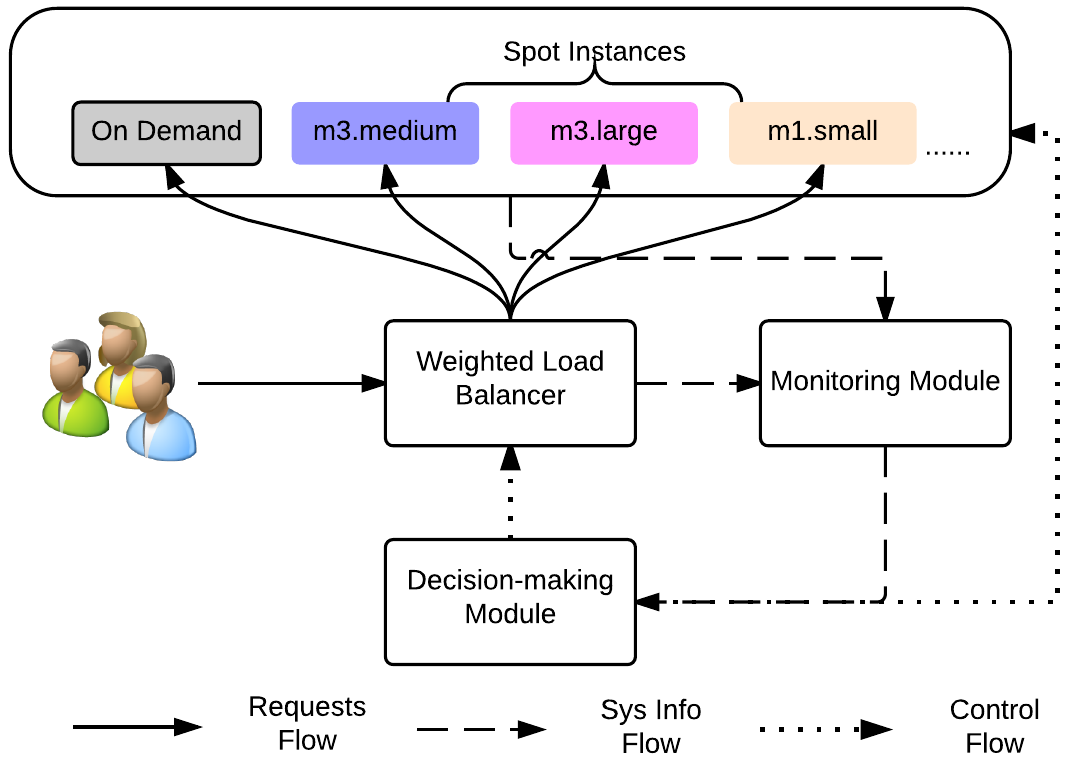}
\caption{Proposed Auto-scaling system architecture}
\label{fig:architecture}
\end{figure}

\begin{table}
\caption{List of Symbols}
\label{tab:symbols}
\begin{center}
\begin{tabular} {c | l}
\hline
\hline
\textbf{Symbol} & \multicolumn{1}{c}{\textbf{Meaning}}\\
\hline
$\mathbf{T}$ & The set of spot types \\
$M_{min}$ & The minimum allowed resource margin of an instance \\
$M_{def}$ & The default resource margin of an instance \\
$Q$ & The quota for each spot group \\
$R$ & The required resource capacity for the current load \\
$F_{max}$ & The maximum allowed fault-tolerant level \\
$f$ & The specified fault-tolerant level \\
$O$ & The minimum percentage of on-demand resources\\
& in the provision\\
$S$ & The maximum number of selected spot groups\\
& in the provision\\
$r_{o}$ & The resource capacity provisioned by on-demand \\
& instances \\
$s$ & The number of chosen spot groups \\
$vm$ & The VM type \\
$vm_{o}$ & The on-demand VM type \\
$c_{vm}$ & The hourly on-demand cost of the $vm$ type instance \\
$num(c, vm)$ & The function returns the number of $vm$ type \\
& instances required to satisfy resource capacity $c$ \\
$C_{o}$ & The hourly cost of provision in on-demand mode \\
$tb_{vm}$ & The truthful bidding price of $vm$ spot group \\
$m$ & The dynamic resource margin of an instance \\
\hline
\end{tabular}
\end{center}
\end{table}

\section{System Model}

For reader's convenience, the symbols used in this paper are listed in Table \ref{tab:symbols}.

\label{sec:system_model}
\subsection{Auto-scaling System Architecture}

As illustrated in Figure \ref{fig:architecture}, our auto-scaling system provisions a single-tier (usually the application server tier) of an application using a mixture of on-demand instances and spot instances. The provisioned on-demand instances are homogeneous instances that are most cost-efficient regarding the application, whilst spot instances are heterogeneous.

Like other auto-scaling systems, our system is composed of the \emph{monitoring} module, the \emph{decision-making} module, and the \emph{load balancer}. The monitoring module consists of multiple independent monitors that are responsible for fetching newest corresponding system information such as resource utilizations, request rates, spot market prices, and VMs' statuses into the system. The decision-making module then makes scaling decisions according to the obtained information based on the predefined strategies and policies when necessary. Since in our proposed system provisioned virtual cluster is heterogeneous, the load balancer should be able to distribute requests according to the capability of each attached VM. The algorithm we use in this case is \emph{weighted round robin}.

The application hosted by the system should be stateless. This restriction does not reduce the applicability of the system as modern cloud applications are meant to de developed in a stateless way in order to realize high scalability and availability \cite{wilder2012}. In addition, stateful applications can be easily transformed into stateless services using various means, e.g., storing the session data in a separated memcache cluster.

\begin{figure}
\begin{center}
\subfigure[]
{\label{fig:naive-a}
\includegraphics[width=0.6in]{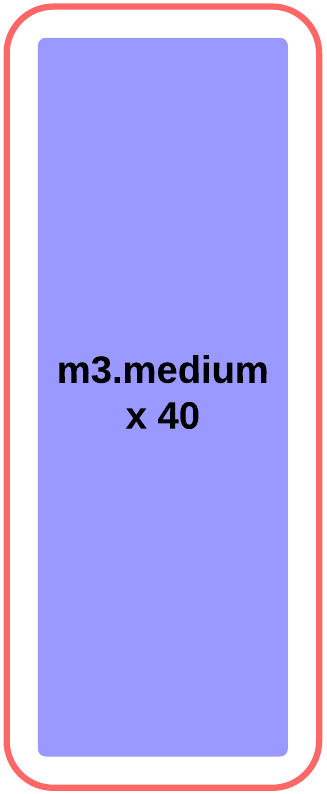}}
\hspace{0.3in}
\subfigure[]
{\label{fig:naive-b}
\includegraphics[width=0.6in]{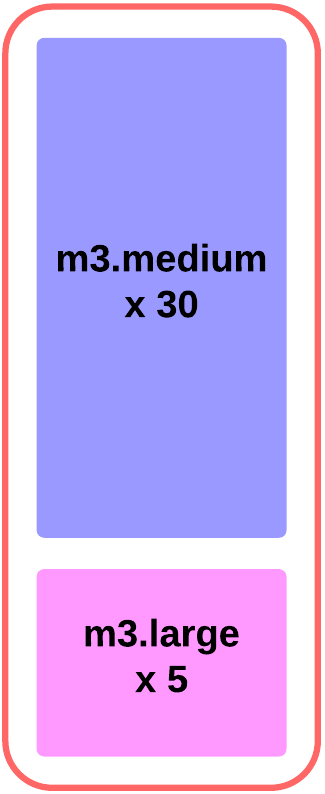}}
\hspace{0.3in}
\subfigure[]
{\label{fig:naive-c}
\includegraphics[width=0.6in]{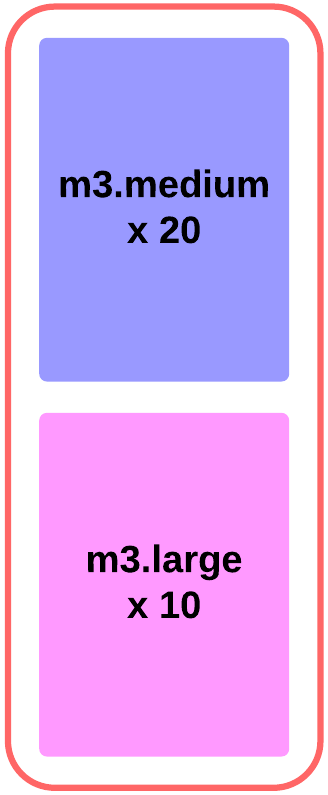}}
\end{center}
\caption{Naive provisioning using spot instances\protect\footnotemark}
\label{fig:naive}
\end{figure}

\footnotetext{The red rectangles in Figure \ref{fig:naive}, \ref{fig:ft}, \ref{fig:ft-2-more}, and \ref{fig:ft-on-demand} stand for the minimum amount of capacity required to process the current workload. Its value is dynamic and proportional to the changing workload so as the amount of redundancy for fault-tolerance.}

\subsection{Fault-Tolerant Mechanism}
Suppose there are sufficient temporal gaps between price variation events of various types of spot VMs, increasing spot heterogeneity in provision can improve robustness. As illustrated in Figure \ref{fig:naive-a}, the application is fully provisioned using 40 \emph{m3.medium} spot VMs only, which may lead it to losing 100\% of its capacity when \emph{m3.medium}'s market price go beyond the bidding price. By respectively provisioning 75\% and 25\% of the total required capacity using 30 \emph{m3.medium} and 5\footnote{According to Amazon's specification, the capacity of 1 m3.large instance is equal to the capacity of 2 m3.medium instances.} \emph{m3.large} spot VMs in Figure \ref{fig:naive-b}, it will lose at most 75\% of its processing capacity when the price of either chosen type rises above the bidding price. Furthermore, if it is provisioned with equal capacity using the two types of spot VMs, like in Figure \ref{fig:naive-c}, termination of the either type of VMs will only cause it to lose 50\% of its capacity.

\begin{figure}
\begin{center}
\subfigure[f-0]
{\label{fig:ft-a}
\includegraphics[width=0.6in]{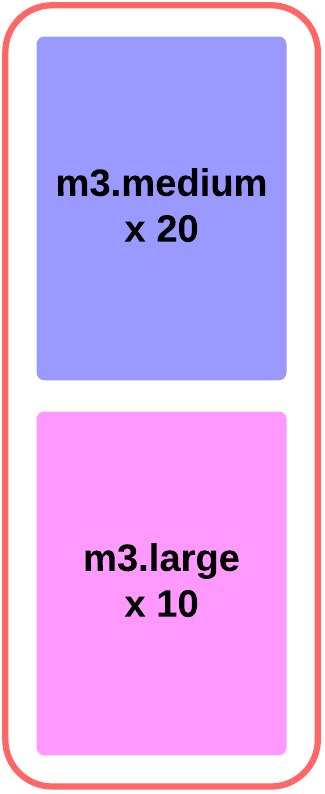}}
\hspace{0.1in}
\subfigure[f-1]
{\label{fig:ft-b}
\includegraphics[width=1.11in]{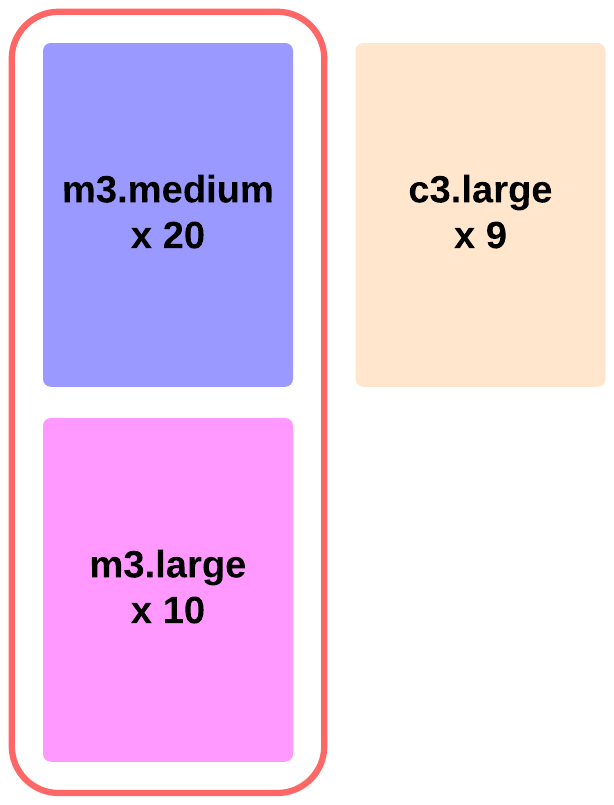}}
\hspace{0.1in}
\subfigure[f-2]
{\label{fig:ft-c}
\includegraphics[width=1.1in]{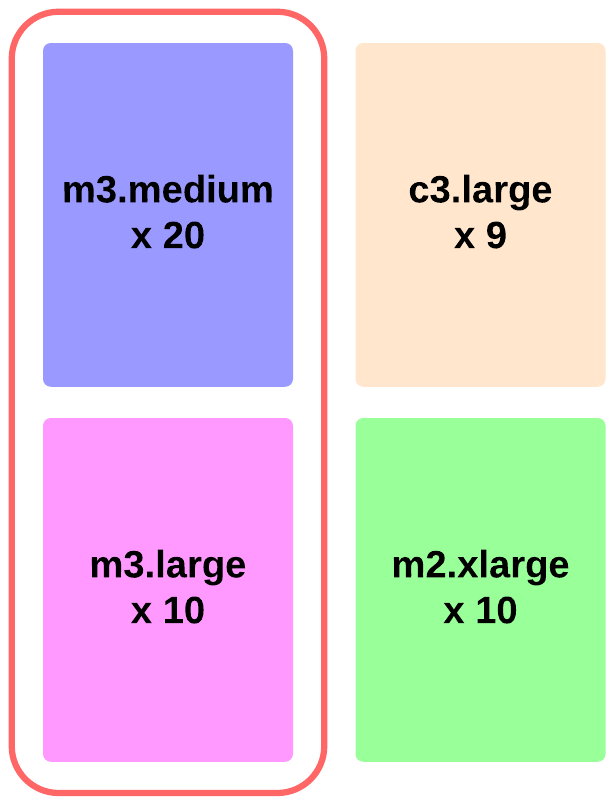}}
\hspace{0.1in}
\subfigure[f-3]
{\label{fig:ft-d}
\includegraphics[width=1.6in]{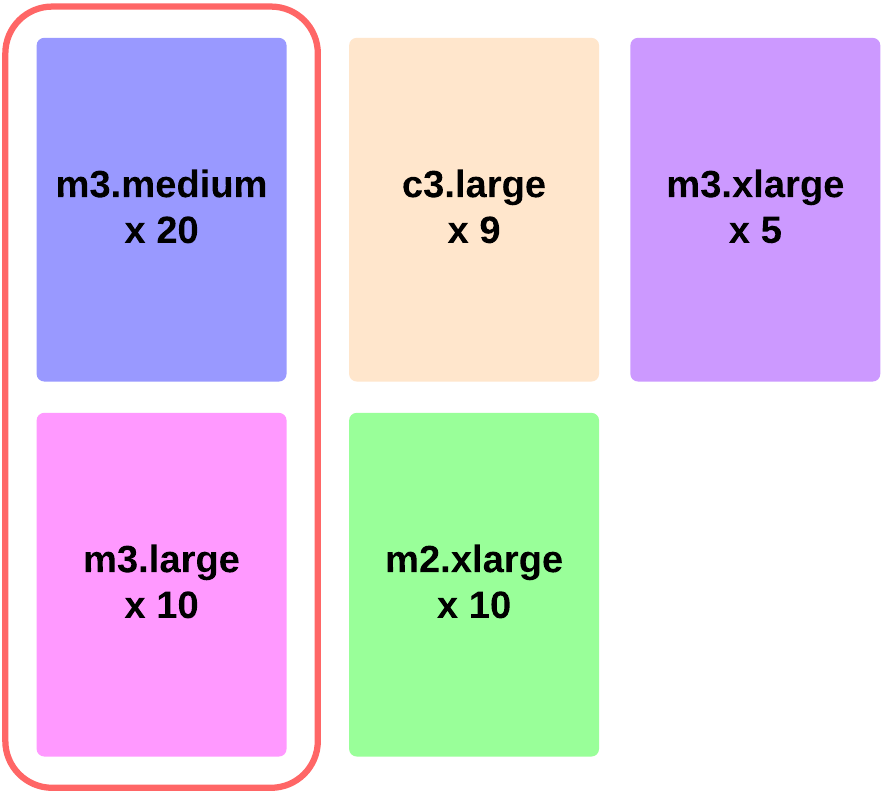}}
\end{center}
\caption{Provisioning for different fault-tolerant levels}
\label{fig:ft}
\end{figure}

This is still unsatisfactory as we demand application performance to be intact even when unexpected termination happens. Simply, the solution is to further over-provision the same amount of capacity using another spot type, as the example illustrated in Figure \ref{fig:ft-b}, it can be 50\% of the required capacity provisioned using 9 \emph{c3.large} instances. In this way, the application is now able to tolerate the termination of any involving type of VMs and remain fully provisioned. After detection of the termination, the scaling system can either provision the application using another type of spot VMs or switch to on-demand instances. Application performance is unlikely to be affected if there is no other termination happens before the scaling operation that repairs the provision fully completes.

However, it takes quite a long time to acquire and boot a VM (around 2 minutes for on-demand instances and 12 minutes for spot instances \cite{Ming}). Hence, there is substantial possibility that another type of spot VMs could be terminated within this time window. To counter such situation, it requires further over-provision the application using extra spot types. We define the \emph{\textbf{fault-tolerant level}} of our auto-scaling system as the maximum number of spot types that can be unexpectedly terminated without affecting application performance before its provision can be fully recovered. Figure \ref{fig:ft} respectively shows the provision examples that comply with fault-tolerant level zero, one, two, and three in our definition with each spot type provisioning 50\% of the required capacity.

Note that setting fault-tolerant level to zero is usually not recommended. Though using multiple types of spot instances confines amount of resource loss when failures happen, with no over-provision to compensate resource loss, it may frequently cause performance degradations as failure probability becomes higher when more types of spot instances are involved.

\subsection{Reliability and Cost Efficiency}
\label{subsec:rel_cost_eff}

\begin{figure}
\begin{center}
\subfigure[f-0]
{\label{fig:ft-a-2-more}
\includegraphics[width=0.6in]{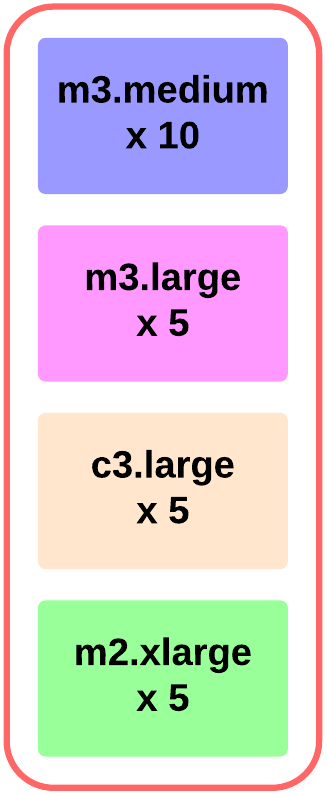}}
\hspace{0.3in}
\subfigure[f-1]
{\label{fig:ft-b-2-more}
\includegraphics[width=1.11in]{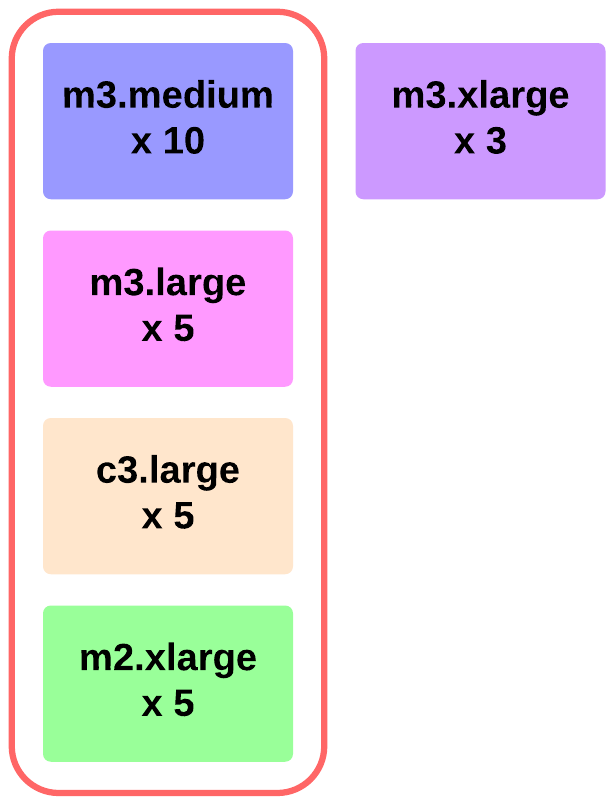}}
\hspace{0.3in}
\subfigure[f-2]
{\label{fig:ft-c-2-more}
\includegraphics[width=1.11in]{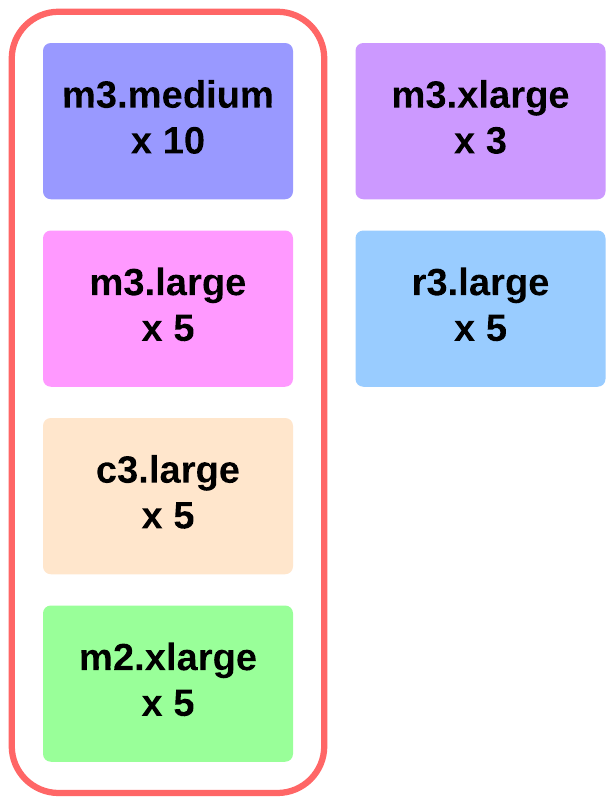}}
\hspace{0.3in}
\subfigure[f-3]
{\label{fig:ft-d-2-more}
\includegraphics[width=1.11in]{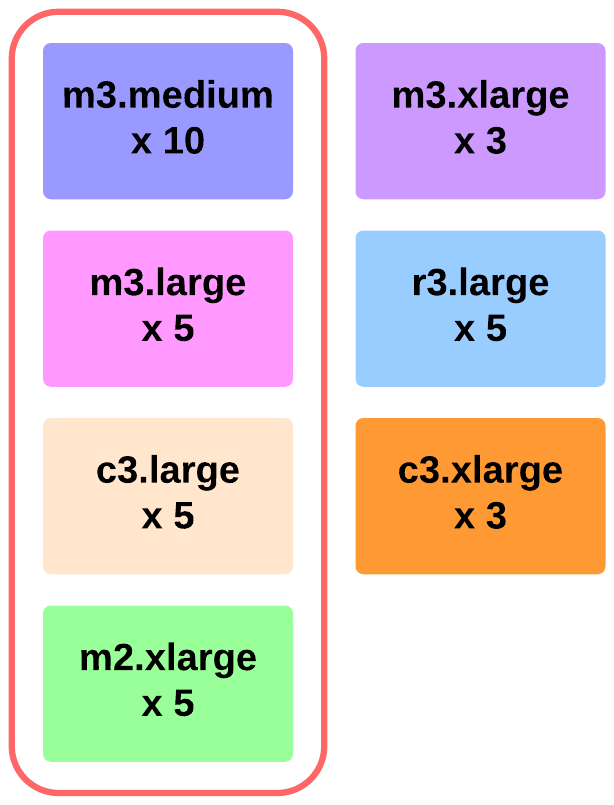}}
\end{center}
\caption{Provisioning for different fault-tolerant levels using 2 more spot types}
\label{fig:ft-2-more}
\end{figure}

Though the provisions shown in Figure \ref{fig:ft-b}, \ref{fig:ft-c}, and \ref{fig:ft-d} successfully increase reliability of the application, they are not cost-efficient. The three provisions respectively over-provision 50\%, 100\%, and 150\% of resources required by the application, which greatly diminishes the cost saving of using spot instances.

One possible improvement is to provision the application using more number of spot types. The illustrative provisions in Figure \ref{fig:ft-2-more} employ two more spot types than that are used in Figure \ref{fig:ft} to reach the corresponding fault-tolerant levels. As the result, total over-provisioned capacities for the three cases are reduced to 25\%, 50\%, and 75\%. Though the provisions now might become more volatile with more types of spot VMs involved, the increased risk is manageable by the fault-tolerant mechanism with over-provision.

\begin{figure}
\begin{center}
\subfigure[f-0]
{\label{fig:ft-a-on-demand}
\includegraphics[width=0.6in]{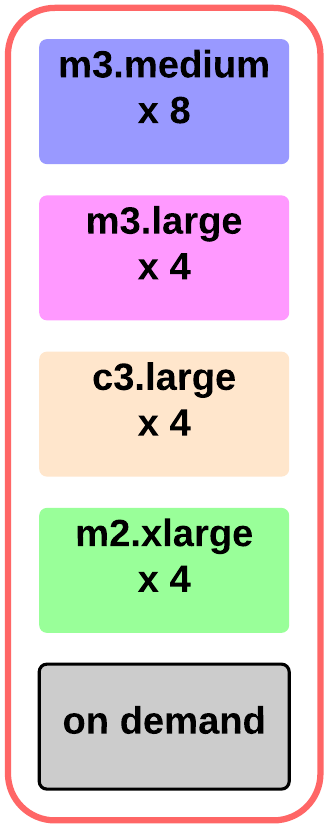}}
\hspace{0.3in}
\subfigure[f-1]
{\label{fig:ft-b-on-demand}
\includegraphics[width=1.11in]{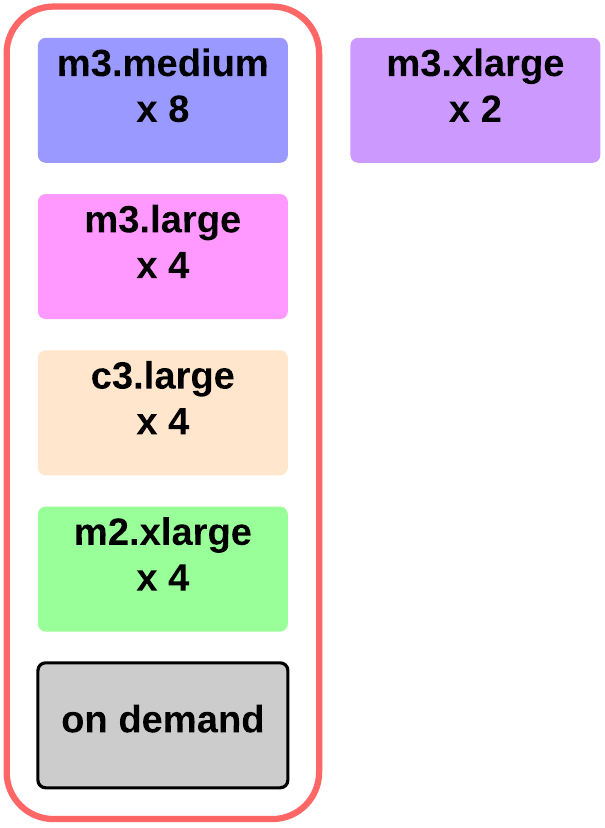}}
\hspace{0.3in}
\subfigure[f-2]
{\label{fig:ft-c-on-demand}
\includegraphics[width=1.11in]{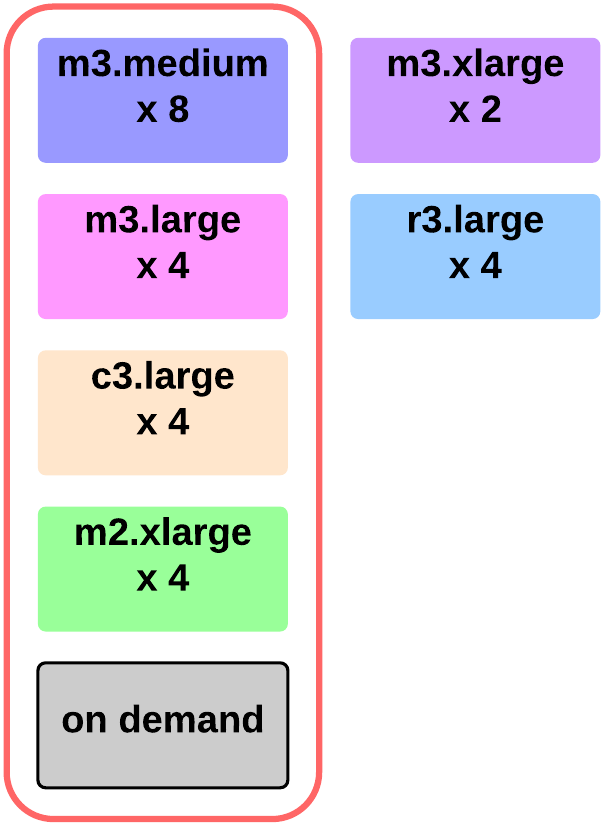}}
\hspace{0.3in}
\subfigure[f-3]
{\label{fig:ft-d-on-demand}
\includegraphics[width=1.11in]{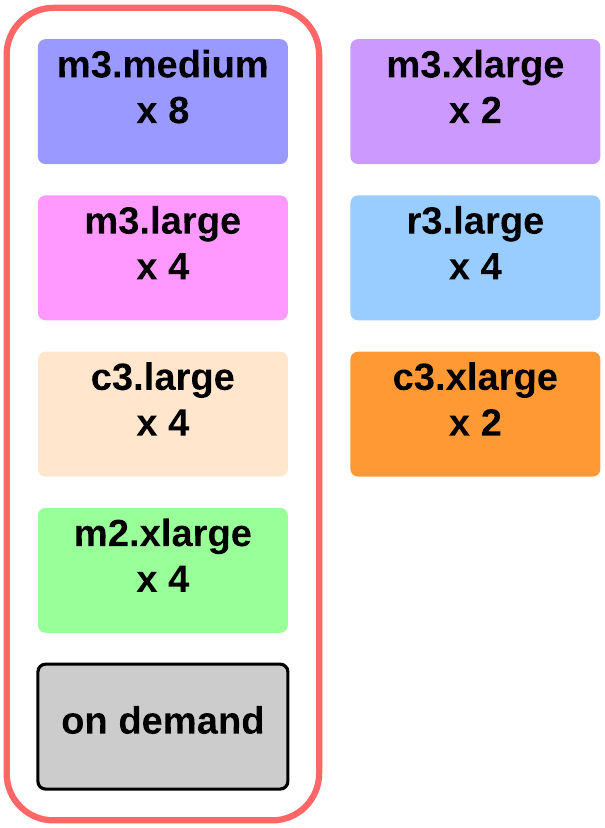}}
\end{center}
\caption{Provisioning for different fault-tolerant levels using mixture of on-demand and spot instances}
\label{fig:ft-on-demand}
\end{figure}

To reduce over-provision, the other choice is to provision the application with a mixture of on-demand instances and spot instances. Like the demonstrations shown in Figure \ref{fig:ft-on-demand}, there are now only 20\%, 40\%, and 60\% over-provisioned capacities if 20\% of the required resource capacity is provisioned by on-demand instances. Moreover, using on-demand resources also further confines amount of capacity that could be lost unexpectedly, thus, improving robustness. On the other hand, this method incurs more financial cost.

We define total capacity that is provisioned by the same type of spot VMs as a \emph{\textbf{Spot Group}}. In addition to that, we give definition to \emph{\textbf{Quota}} ($Q$), which is the capacity each spot group needs to provision given the capacity provisioned by on-demand resources ($r_{o}$) and the fault-tolerant level ($f$). It is calculated as:

\begin{equation}
\label{eq:quota}
Q = \frac{R - r_{o}}{s - f}
\end{equation}

\noindent where $R$ represents the required capacity for the current load, and $s$ denotes the number of chosen spot types. The minimum amount of capacity that is required to over-provision then can be calculated as $Q * f$.

We call a provision is \emph{\textbf{safe}} if the provisioned capacity of each spot group is larger than $Q$. Hence, the problem of scaling web applications using heterogeneous spot VMs is transformed to dynamically selecting spot VM types and provisioning corresponding spot and on-demand VMs to keep the provision in safe state with minimum cost when the application workload increases, and timely deprovisioning various types of VMs when they are no longer needed.

\section{Scaling Policies}
\label{sec:scaling_policies}

Based on the previous fault-tolerant model, we propose cost-efficient auto-scaling policies that comply with the defined fault-tolerant semantics for hourly billed cloud market like Amazon EC2.

\subsection{Capacity Estimation and Load Balancing}
Our auto-scaling system is aware of multiple resource dimensions (such as CPU, Memory, Network, and Disk I/O). It needs the profile of the target application regarding its average resource consumption for all the considered dimensions. Currently, the profiling needs to be performed offline, but our approach is open to integrate dynamic online profiling into it. 

With the profile, the system is able to estimate the processing capability of each spot type under the context of the scaling application. Based on that, it can easily determine how to distribute incoming requests to the heterogeneous VMs to balance their loads. 
In addistion, the estimated capabilities are used in the calculation of scaling plans as well.

\subsection{Spot Mode and On-Demand Mode}
Our scaling system runs interchangeably in \emph{\textbf{Spot Mode}} and \emph{\textbf{On-Demand Mode}}. Spot Mode provisions application in the way explained in Section \ref{subsec:rel_cost_eff}. In Spot Mode, user needs to specify the minimum percentage of required resources provisioned by on-demand instances, symbolized as $O$. He can also set a limit on the number of selected spot groups in provision, denoted as $S$. To define these parameters, users can utilize the simulation tool implemented by us (described in Section \ref{sec:implementation}) to find the optimal configurations according to the recent spot market history without running real tests on the cloud. Furthermore, these parameters can be dynamically adjusted using machine learning technologies. We leave this as our future work. In On-Demand Mode, application is fully provisioned by on-demand instances without over-provision. Switches between modes are dynamically triggered by the scaling policies detailed in the following sections.

\subsection{Truthful Bidding Prices}
Bidding truthfully means the participant in an auction always bids the maximum price he is willing to pay. In order to guarantee cost-efficiency, truthful bidding price for each VM type in our policies is calculated dynamically according to real-time workload and provision. Before computing them, we first calculate the hourly baseline cost if the application is provisioned in On-Demand Mode, which can be represented as:


\begin{equation}
\label{eq:on_demand_price}
C_{o} = num(R, vm_{o}) * c_{vm_{o}}
\end{equation}

\noindent where function $num(R, vm_{o})$ returns the minimum number of instances of on-demand VM type required to process the current workload. $c_{vm_{o}}$ is the on-demand hourly price of on-demand instance type. Then truthful bidding price of spot type $vm$ is derived as follow:


\begin{equation}
\label{eq:tb_price}
tb_{vm} = \frac{C_{o} - num(r_{o}, vm_{o}) * c_{vm_{o}}}{s * num(Q, vm)}
\end{equation}

\noindent where $num(r_{o}, vm_{o})$ and $num(Q, vm)$ are interpreted similarly to $num(R, vm_{o})$ in Equation \eqref{eq:on_demand_price}.

This ensures that even in the worst situation that all chosen spot types' market prices are equal to their corresponding truthful bidding prices, the total hourly cost of the provision will not exceed that in On-Demand Mode.

\begin{algorithm}[!t]
  \caption{Find new provision when the system needs to scale up}
  \label{al:scaling_up_policy}
  \KwIn{$R$ : the current workload}
  \KwIn{$n_{c}$ : the number of on-demand VMs in current provision}
  \KwIn{$vm_{o}$ : the on demand $vm$ type}
  \KwIn{$O$ : the minimum percentage of on-demand resources}
  \KwOut{$target\_provision$}  
  
  $min\_vm_{o} \leftarrow \textbf{max}(n_{c}, num(R * O, vm_{o}))$\; \label{line:min_on_demand}
  $max\_vm_{o} \leftarrow num(R, vm_{o})$\;
  
  $candidate\_set \leftarrow$ call Algorithm \ref{al:search_provison_with_num_on_demand} for each integer $n$ in $[min\_vm_{o}, max\_vm_{o}]$\;
  \Return on-demand provision if $candidate\_set$ \rm is empty\\
  otherwise the provision with minimum cost in $candidate\_set$\; \label{line:switch_to_on_demand}
\end{algorithm}

\begin{algorithm}[!t]
  \caption{Find provision given the number of on-demand instances}
  \label{al:search_provison_with_num_on_demand}
  \KwIn{$n$ : the number of on-demand VMs}
  \KwIn{$g_{c}$ : the set of spot groups in current provision}
  \KwIn{$vm_{o}$ : the on-demand $vm$ type}
  \KwIn{$f$ : the fault-tolerant level}
  \KwIn{$\mathbf{T}$ : the set of spot types}
  \KwIn{$S$ : the maximum number of chosen spot groups}
  \KwOut{$new\_provision$}
  
  $min\_groups \leftarrow \textbf{max}(|g_{c}|, f + 1)$\;
  
  $max\_groups \leftarrow \textbf{min}(|\mathbf{T}|, S)$\;
  
  \If{$max\_groups < min\_groups$}
  {
  	provision not found\;
  }
  \Else {
    \For {$s$ \textbf{from} $min\_groups$ \textbf{to} $max\_groups$}
    {
    	$p \leftarrow p \cup (vm_{o}, n)$\;
  		compute $Q$ using Equation \eqref{eq:quota}\;
  		compute $tb_{vm}$ for each $vm$ in $\textbf{T}$\;
  		$p \leftarrow p \cup g_{c}$\;\label{line:chosen_groups}
  		$groups \leftarrow$ each $group$ not in $g_{o}$ and whose $tb_{vm}$ is higher than market price\;
  		$k \leftarrow s - |g_{c}|$\;
  		\If{$|groups| \geq k$}
  		{
  			$p \leftarrow p \ \cup$ top k cheapest groups in $groups$\;\label{line:cost_efficiency}
  			$provisions \leftarrow provisions \cup p$\;
  		}
    }
  }
  \Return the cheapest provision in $provisions$\;
\end{algorithm}

\begin{algorithm}[!t]
  \caption{Find target provision when the billing hour of one on-demand instance is about to end}
  \label{al:scaling_down_on_demand}
  \KwIn{$R$ : the current workload}
  \KwIn{$n_{c}$ : the number of on-demand instances in current provision}
  \KwIn{$vm_{o}$ : the on-demand $vm$ type}
  \KwIn{$O$ : the minimum percentage of on-demand resources}
  \KwOut{$target\_provision$}
  \If{$n_{c} \leq num(R * O, vm_{o})$}
  {\label{line:check_on_demand_sufficiency_start}
  	provision not found\;\label{line:check_on_demand_sufficiency_end}
  } 
  \Else {
  	$p_{1} \leftarrow $ call Algorithm \ref{al:search_provison_with_num_on_demand} with $n_{c}$\; \label{line:provision_with_the_instance}
  	$p_{2} \leftarrow$ call Algorithm \ref{al:search_provison_with_num_on_demand} with $n_{c} - 1$\; \label{line:provision_without_the_instance}
  	\Return on-demand provision if neither $p_{1}$ nor $p_{2}$ is found
  	otherwise either provision that is cheaper\; \label{line:provision_with_less_cost}
  }
  
\end{algorithm}

  

\subsection{Scaling Up Policy}
Scaling up policy is called when some instances are terminated unexpectedly or the current provision cannot satisfy resource requirement of the application. By resource requirement, in Spot Mode, it means the provision should be \emph{\textbf{safe}} under the current workload, which is defined in Section \ref{subsec:rel_cost_eff}. While in On-Demand Mode, it only requires the resource capacity of the provision to exceed the resource needs of the current workload.

Algorithm \ref{al:scaling_up_policy} is used to find the ideal new provision when the system needs to scale up. To avoid frequent drastic changes, the algorithm only provisions VMs incrementally. As shown by line \ref{line:min_on_demand} in Algorithm \ref{al:scaling_up_policy}, it limits the number of provisioned on-demand instances to be at least its current number. For each valid number of on-demand instances, it calls Algorithm \ref{al:search_provison_with_num_on_demand} to find the corresponding best provision among provisions with various combinations of spot groups. Similarly, in Algorithm \ref{al:search_provison_with_num_on_demand} (line \ref{line:chosen_groups}), it retains the spot groups chosen by the current provision and only incrementally adds new groups according to their cost-efficiency (line \ref{line:cost_efficiency}). If there is no valid provision found, the system switches to on-demand mode.

After the target provision is found, the system compares it with the current provision and then contacts the cloud provider through its API to provision the corresponding types of VMs that are in short.

In the worst case, the time complexity of the scaling up policy is $O(N*S*|\textbf{T}|))$ where $N$ is the number of on demand instances required to provision the current workload in on demand mode, $S$ denotes the maximum number of chosen spot groups, and $|\textbf{T}|$ is the number of spot types considered. Since the parameters are all small integers, the computation overhead of the algorithm is acceptable in an online decision making scenario.

\subsection{Scaling Down Policy}
Since each instance is billed hourly, it is unwise to shut down one instance before its current billing hour matures. We therefore put the decision of whether each instance should be terminated or not at the end of their billing hours. The specific decision algorithms are different for on-demand instances and spot instances.

\subsubsection{Policy for on-demand instances}
When one on-demand instance is at the end of its billing hour, we not only need to decide whether the instance should be shut down, but also have to make changes to the spot groups if necessary. The summarized policy is abstracted in Algorithm \ref{al:scaling_down_on_demand}. The algorithm first checks whether enough on-demand instances are provisioned to satisfy the on-demand capacity limit (line \ref{line:check_on_demand_sufficiency_start} and line \ref{line:check_on_demand_sufficiency_end}). If there are sufficient on-demand instances, it endeavours to find the most cost-efficient provisions with and without the on-demand instance by calling Algorithm \ref{al:search_provison_with_num_on_demand} (line \ref{line:provision_with_the_instance} and line \ref{line:provision_without_the_instance}). Suppose the current provision is in On-Demand Mode and no provision is found without the on-demand instance, the provision will remain in On-Demand Mode. Otherwise, if a new provision is found without the current instance, the policy switches the provision to Spot Mode. In the case that the current provision is already in Spot Mode, it picks whichever provision that incurs lower hourly cost.

\subsubsection{Policy for spot instances}
When dealing with a spot instance whose billing period is ending, in the base policy, we simply shut down the instance when the corresponding spot quota $Q$ can be satisfied without it. Thereafter, the policy will evolve with the introduced optimizations in Section \ref{sec:optimizations}.

\subsection{Spot Groups Removal Policy}
Note that in both scaling up and down policies, we forbid removing selected spot groups from provision. Instead, we evict a chosen spot group when any spot instances of such type is terminated by the provider. Since bidding price of each instance is calculated dynamically, instances within the same spot group may be bid at different prices. This could cause some instances to remain alive even after the corresponding spot groups are removed from provision. We call the instances that are running but do not belong to any group \emph{\textbf{orphans}}. Though orphan instances are still in production, they are not considered a part of the provision according to the fault-tolerant semantics when making scaling decisions. In the base policies, although they will not be shut down until their billing hour ends, extra instances still need to be launched to comply with the fault-tolerant semantics, which causes resource waste. This drawback is addressed by the introduced optimizations in the following section.

\section{Optimizations}
\label{sec:optimizations}

We have made several optimizations on the above proposed base policies to further improve cost-efficiency and reliability of the system.

\subsection{Bidding Strategy}

In the scaling policies, spot groups are bid at truthful bidding prices calculated by Equation \eqref{eq:tb_price} due to cost-efficiency concern. While focusing on robustness, the system can employ a different strategy to bid higher so as to grasp spot instances as long as possible.

\subsubsection{Actual Bidding Strategies}
There are two actual bidding strategies, namely truthful bidding strategy and on-demand price bidding strategy embedded in the system.

\begin{itemize}
\item \textbf{Truthful Bidding Strategy:} the system always bids the truthful bidding price calculated by Equation \eqref{eq:tb_price} when new spot instances are launched. Since partial billing hours ended by cloud provider are free of charge, cloud users can save money by letting cloud provider terminate their spot instances once their market prices exceed the corresponding truthful bidding prices. On the other hand, it leads to more unexpected terminations.

\item \textbf{On-Demand Price Bidding Strategy:} the system always bids the on-demand price of the corresponding spot type whenever trying to obtain new spot instances. This strategy will cost cloud users more money but provides a higher level of protection against unexpected terminations.
\end{itemize}

\subsubsection{Revised Spot Groups Removal Policy}
In the base policies, less cost-efficient spot groups could remain in provision for a long time unless some of their instances are terminated by provider. When the actual bids are higher than the truthful bidding prices, the situation could become worse. Instead of just relying on provider terminating uneconomical spot groups, the revised policy actively inspects whether market prices of some spot groups have exceeded their corresponding truthful bidding prices and remove them from the provision. In the meantime, for spot groups whose market prices are still below their truthful bidding prices, it looks for chance to replace them by more economical spot groups that have not been selected. To minimize disturbance to provision, such operations should be conducted in a long interval, such as every 30 minutes in our implementation. Members of removed or replaced spot groups become orphans.

\begin{figure}
\begin{center}
\subfigure[launching 2 new m1.small instances for m1.small spot group]
{\label{fig:util-orphan-a}
\includegraphics[width=4in]{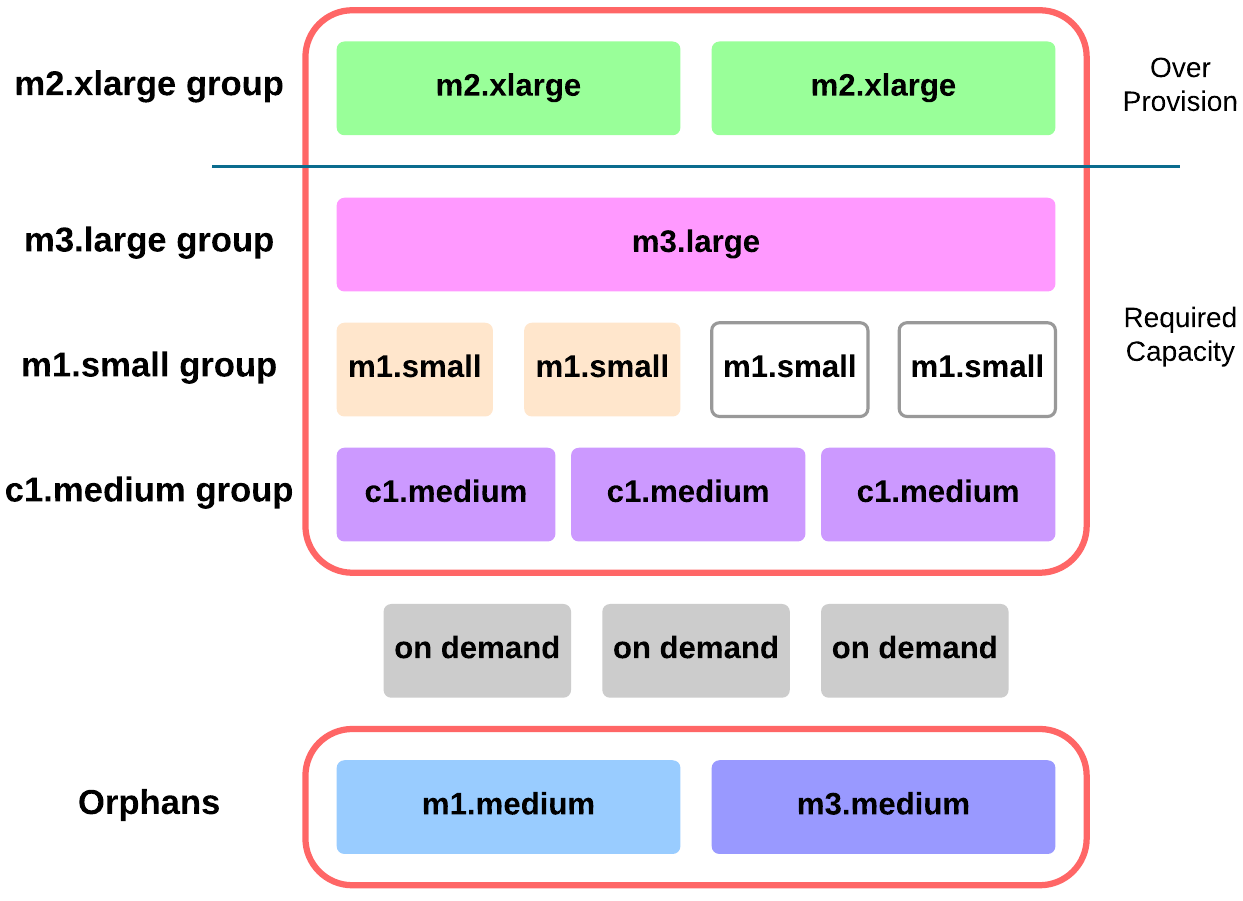}}
\subfigure[using one m1.medium orphan to temporarily substitute 2 m1.small instances]
{\label{fig:util-orphan-b}
\includegraphics[width=4in]{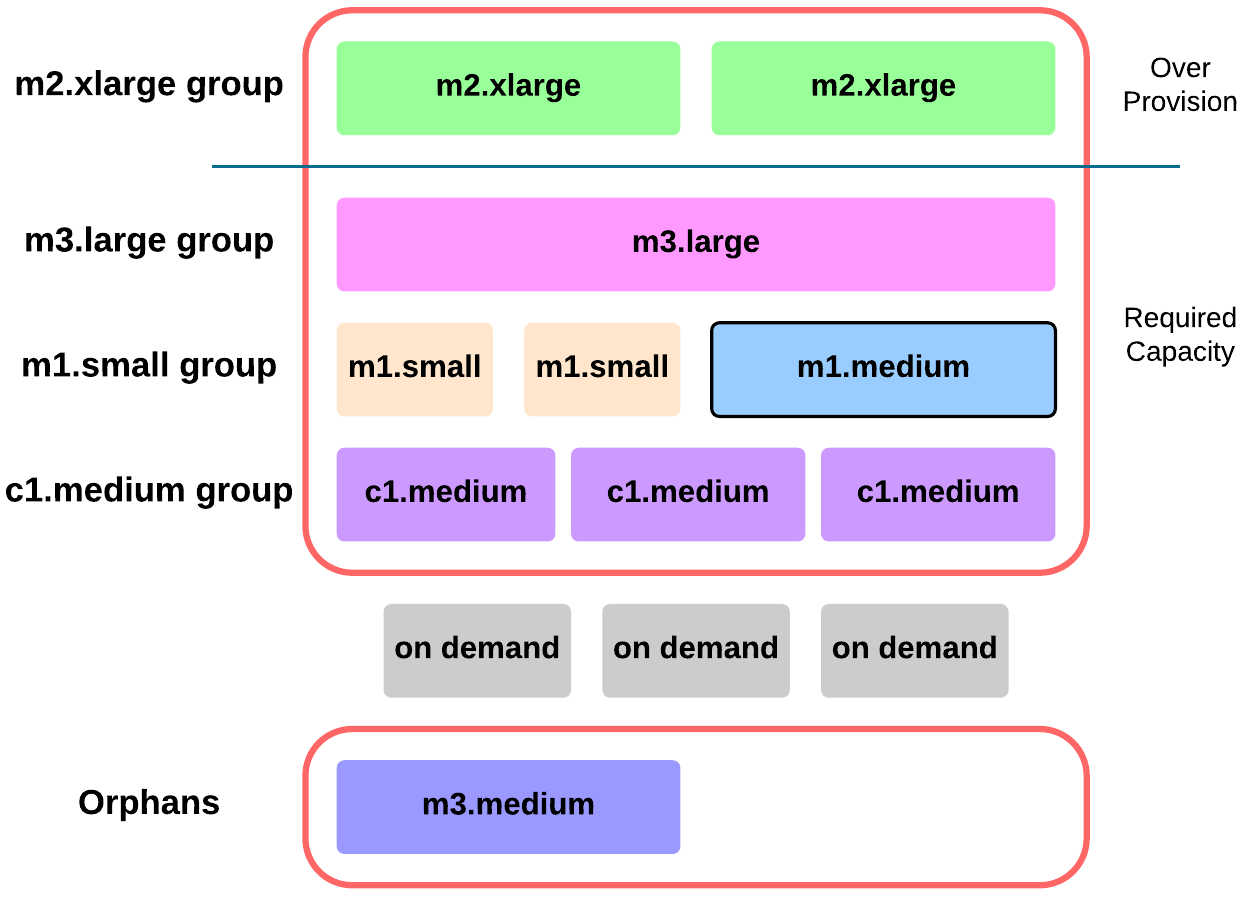}}
\end{center}
\caption{Provisioning with orphans under fault-tolerant level one}
\label{fig:util-orphan}
\end{figure}

\subsection{Utilizing Orphans}
After removing or replacing some spot groups, if the system simply lets members of these spot groups become orphans and immediately start instances of newly chosen spot groups, the stability of provision will be affected. Furthermore, as orphans are not considered as valid capacity in the base polices, during the transition period, it has to provision more resources than necessary, which results in monetary waste.

To alleviate this problem, we aim to utilize as many orphans in provision as possible to deter the time to provision new VMs. As a result, resource waste can be reduced and cost-efficiency is improved.

We modify the proposed fault-tolerant model to allow a spot group temporarily accept instances that are heterogeneous to the spot group type under certain conditions. Figure \ref{fig:util-orphan} illustrates such provision. In Figure \ref{fig:util-orphan-a}, the \emph{m1.small} group does not have sufficient instances to satisfy its quota. Instead of launching 2 new \emph{m1.small} spot instances, the policy now temporarily move the available orphan, one \emph{m1.medium} instance, to \emph{m1.small} group to compensate the deficiency of its quota. Even though \emph{m1.small} group becomes heterogeneous in this case, it does not violate the fault-tolerant semantics as losing any type of spot instances will not influence the application performance. However, in some situations, heterogeneity in spot groups could cause violation of the fault-tolerant semantics, for example, there might be case that three \emph{m1.medium} orphans are spread across three spot groups and the total capacity of the three instances exceeds the spot quota. Then losing the three \emph{m1.medium} instances will violate the fault-tolerant semantics. Fortunately, such cases are very rare as orphans are usually small in numbers and are expected to be shut down in a short time.

With this relaxation of the fault-tolerant model, the previous scaling up and scaling down policies need to be revised to efficiently utilize capacity of orphans.

\subsubsection{Revised Scaling Up Policy}
The new scaling up policy uses the same algorithm (Algorithm \ref{al:scaling_up_policy}) to find the target provision. However, instead of simply launching instances to reach the target provision, the new policies take a deeper thought whether it can utilize existing orphans to meet the quota requirements in the target provision.

The new policy first checks whether the target provision chooses new spot groups. If there are orphans whose types are the same to any newly chosen groups, lying either within orphan queue or other spot groups, they are immediately moved to the corresponding new spot groups. After that, the policies endeavour to insert non-utilized orphans from the orphan queue into spot groups that have not met their quota requirement. If all the orphans have been utilized and some groups still cannot satisfy their quota, new spot instances of the corresponding types then will be launched.

\subsubsection{Revised Scaling Down Policy}
Regarding policy for on-demand instances that are close to their billing hour, the new policy utilizes the same mechanism in the revised scaling up policy to provision any changes between the current provision and the target provision.

For spot scaling down policy, if the spot instance is in orphan queue, it is immediately shut down. Suppose it is within the spot group of the same type, it is shut down when the spot quota can be satisfied without it. In the case that the instance is an orphan within other spot group, the new policy shuts down the instance and in the meantime starts certain number of spot instances of the spot group type to compensate the capacity loss.

\subsection{Reducing Resource Margin}
For applications running on traditional auto-scaling platform, administrator usually leaves a margin at each instance to handle short-term workload surge in order to buy time for booting up new instances. This margin empirically ranges from 20 to 25\% of the instance's capacity.

With over-provision already in place in our system, this margin can be reduced under Spot Mode provision. We devise a mechanism that dynamically changes the margin according to the current fault-tolerant level. Since higher fault-tolerant level leads to more over-provision, we can be more aggressive in reducing the margin of each instance. In detail, the dynamic margin is determined by the formula:

\begin{equation}
\begin{split}
m = \frac{M_{def} - M_{min}}{F_{max}} * f + M_{min}
\end{split}
\end{equation}

\noindent where $M_{min}$ means the minimum allowed margin, e.g., $10\%$, $M_{def}$ is the default margin used without dynamic margin reduction, e.g., $25\%$, and $F_{max}$ is the maximum allowed fault-tolerant level.

\begin{figure}[!t]
\centering
\includegraphics[width=5in]{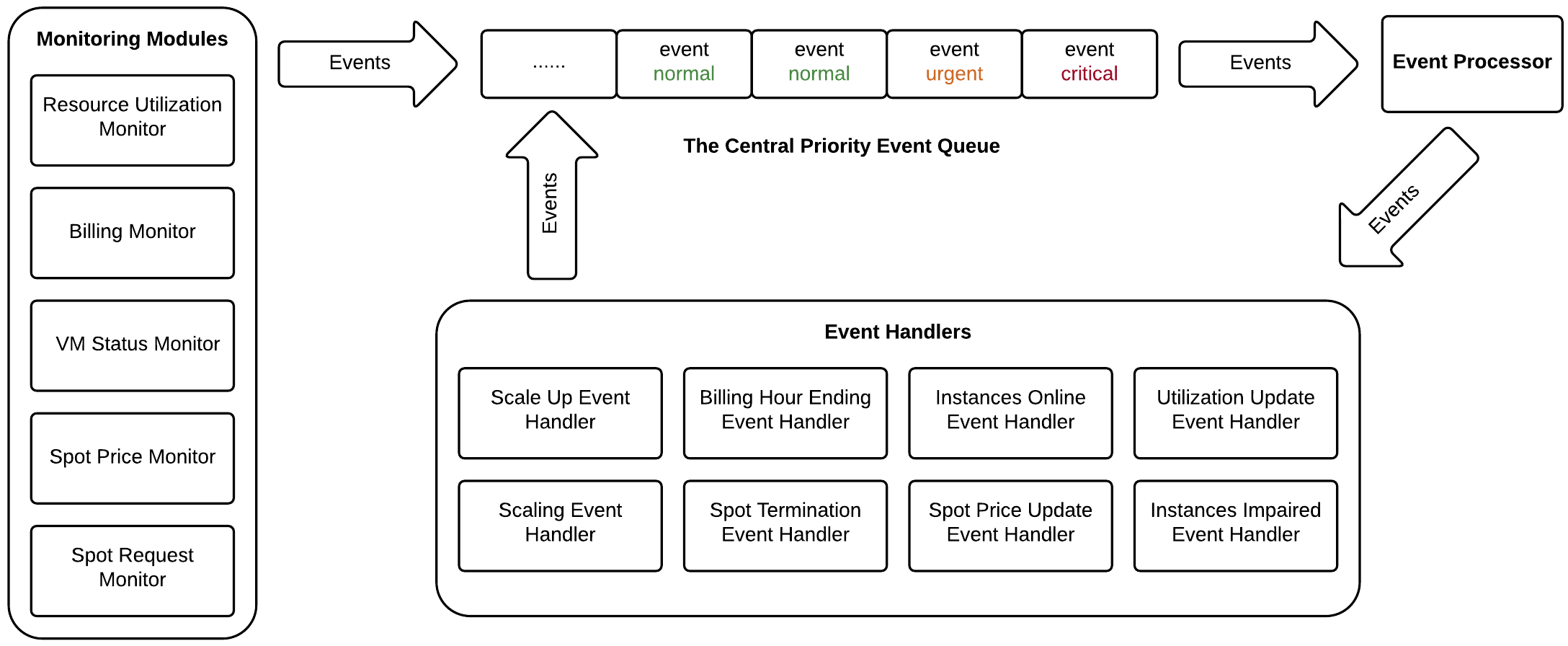}
\caption{Components of the Implemented Auto-scaling System}
\label{fig:implementation}
\end{figure}
\section{Implementation}
\label{sec:implementation}

We implemented a prototype of the proposed auto-scaling system on Amazon EC2 platform using Java, the components of which are illustrated in Figure \ref{fig:implementation}. It employs an event-driven architecture with the monitoring modules continuously generating events according to newly obtained information, and the central processor consuming events one by one. Monitoring modules produce and insert corresponding events with various critical levels into the central priority event queue. They include the \emph{resource utilization monitors} that watch all dimensions of resource consumption of running instances, the \emph{billing monitor} that gazes billing hour of each requested VM, the \emph{VM status monitor} that reminds the system when some instances are online or offline, the \emph{spot price monitor} that records newest spot market prices for each considered spot type, and \emph{the spot request monitor} that surveillances any unexpected spot termination. On the other side, the central event processor fetches events from the event queue and assigns them to the corresponding event handlers that realize the proposed policies to make scaling decisions or perform scaling actions.

The prototype implementation provides a general interface for users to plug different load balancer solutions into the auto-scaling system. In our case, we use \emph{HAProxy} with weighted round robin algorithm. It also offers the interface to allow users to automatically customize configurations of VMs according to their own available resources after they have been booted.

For quick concept validation and repeatable evaluation of the proposed auto-scaling policies, we created a simulation version of the system. The same code base is transplanted onto CloudSim \cite{Calheiros2011} toolkit which provides the underlying simulated cloud environment. Assuming bids from user impose negligible influence on market prices, the simulation tool is able to provide quick and economical validation of the proposed polices using historical data of the application and the spot market as input. 

For more details about the implementation, please refer to the released code\footnote{\url{https://github.com/quchenhao/spot-auto-scaling}}.

\section{Performance Evaluation}
\label{sec:performance_evalutaion}

\subsection{Simulation Experiments}
As stated in Section \ref{sec:implementation}, to allow repeatable evaluation, we developed a simulation version of the system that allows us to compare the performances of different configurations and policies using traces from real applications and spot markets.

\begin{figure}[!t]
\centering
\includegraphics[width=4in]{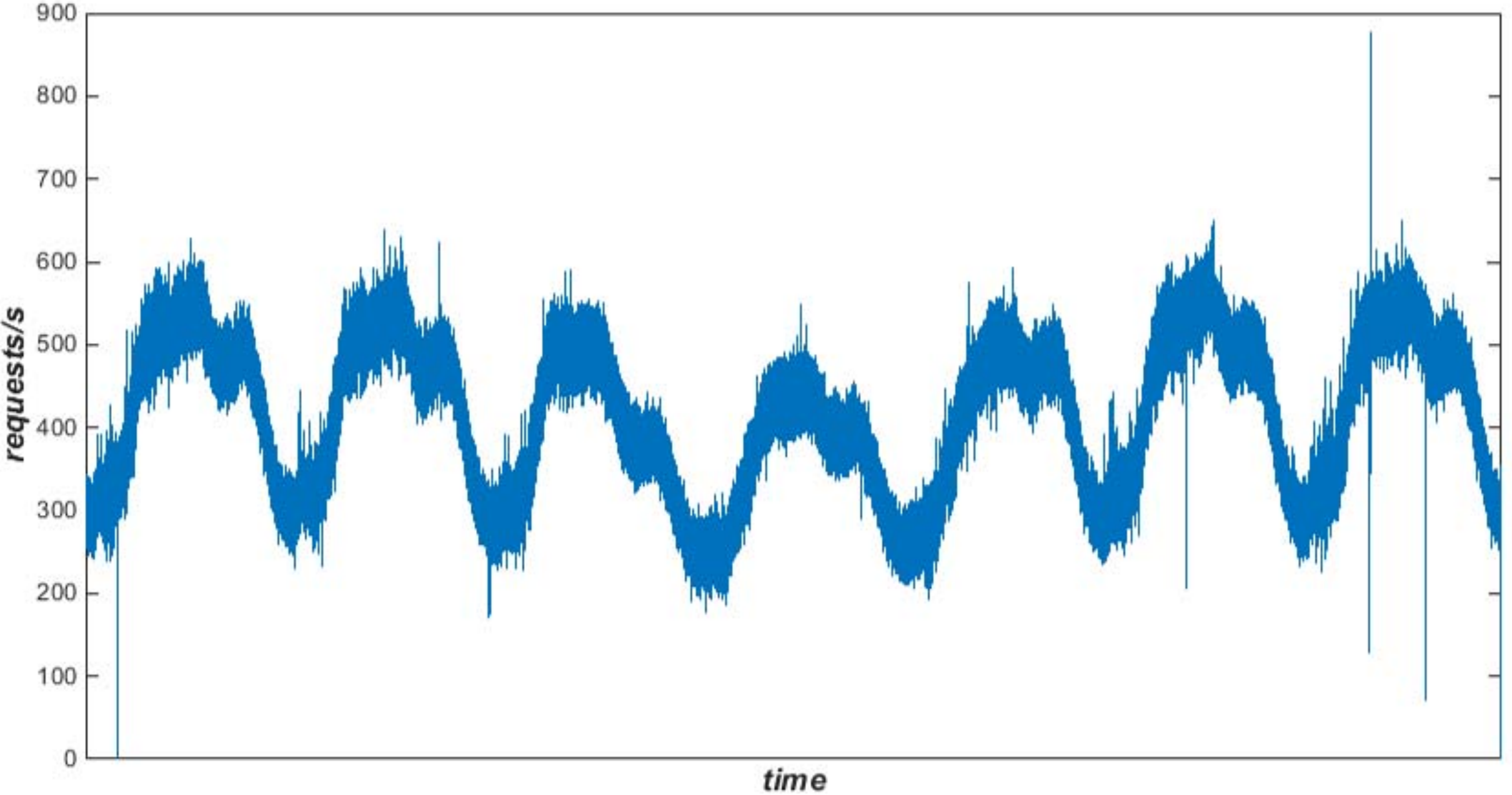}
\caption{The English Wikipedia workload from Sep 19th 2009 to Sep 26th 2009}
\label{fig:workload}
\end{figure}

\subsubsection{Simulation Settings}
We use one week trace of 10\% English Wikipedia requests from Sep 19th 2007 to Sep 26th 2007 as the workload \cite{Baaren2015, Urdaneta2009}, which is depicted in Figure \ref{fig:workload}. Note that our approach is general purpose and can be applied to any workload, as the proposed system does not make assumptions on the workload and is fully reactive. We adopt the Wikipedia workload in experiments because it reveals significant variations that can trigger frequent scaling operations to let us observe the behaviour of our system. We believe one week trace is enough for the purpose of our experiments, as it gives the system ample opportunities to exercise the scaling policies. In addition, as reported by Eldin et al. \cite{Eldin}, the Wikipedia workload revealed strong weekly pattern with only gradual changes in amplitude, level, and shapes.

We consider 13 spot types in Amazon EC2. Their spot prices are simulated according to one week Amazon's spot prices history from March 2nd 2015 18:00:00 GMT in the relatively busy \emph{\textbf{us-east}} region. The involving spot types and their corresponding history market prices are illustrated in Figure \ref{fig:spot_price_history}.

We set requests timeout at 30 seconds. In addition, we respectively set minimum allowed resource margin ($M_{min}$) and default resource margin ($M_{def}$) at 10\% and 25\%. We found out that \emph{c3.large} instance is the most cost-efficient type for the wikipedia application based on a small scale resource profiling test of the Wikibench application \cite{Baaren2009} on Amazon EC2 and the resource specifications of each instance type released by Amazon. It is selected to provision all the on-demand resources in the experiments. All simulation experiments start with 5 \emph{c3.large} on-demand instances. Length of simulated requests are generated following a pseudo Gaussian distribution\footnote{Since Wikipedia is serving mostly the same type of requests - page view, the time taken to process each request is also likely to fall in a certain interval. To coarsely model such behaviour, we utilize Gaussian distribution. Other distributions with small head and tail can serve the same purpose as well.} with mean of 0.07 ECU\footnote{It means the request takes 70ms to finish if it is computed by the VM equipped with vCPU as powerful as 1 Elastic Computing Unit (ECU)} and standard deviation of 0.005 ECU so that different tests using the same random seed are receiving exactly the same workload. The VM start up, shut down, and spot requesting delays are generated in the same way using pseudo Gaussian distribution. The means of the above three distributions are respectively 100, 100, 550 seconds, and the standard deviations are set at 20, 20, 50 seconds. The test results are deterministic and repeatable on the same machine.

We tested our scaling policies with various fault-tolerant levels and different least amounts of on-demand resources, which are represented respectively as ``$f-x$'' and ``$y\%$ on-demand'' in the results. We also tested the polices using the two embedded bidding strategies and static/dynamic resource margins.

We concentrate on two metrics, real-time response time of requests (average response time per second reported) and total cost of instances, in all the experiments.

\subsubsection{Benchmarks}
We compare our scaling policies with two benchmarks:

\begin{itemize}
\item \textbf{On-Demand Auto-scaling:} This benchmark only utilizes on-demand instances. It is implemented by restricting the auto-scaling system always in On-Demand Mode.

\item \textbf{One Spot Type Auto-scaling:} The auto-scaling policies used in this benchmark, like the proposed policies, provision a mixture of on-demand resources and spot resources. The benchmark also has a limit on minimum amount of on-demand resources provisioned. However, for spot instances, it only provisions one spot group that is the most cost-efficient at the moment without over-provision. If the provisioned spot instances are terminated, a new spot group then is selected and provisioned. Suppose a more economic spot group is found, the old spot group is gradually replaced by the new one. It is implemented by setting fault-tolerant level to zero and limiting at most one spot group can be provisioned.
\end{itemize}

\begin{figure}
\centering
\includegraphics[width=4in]{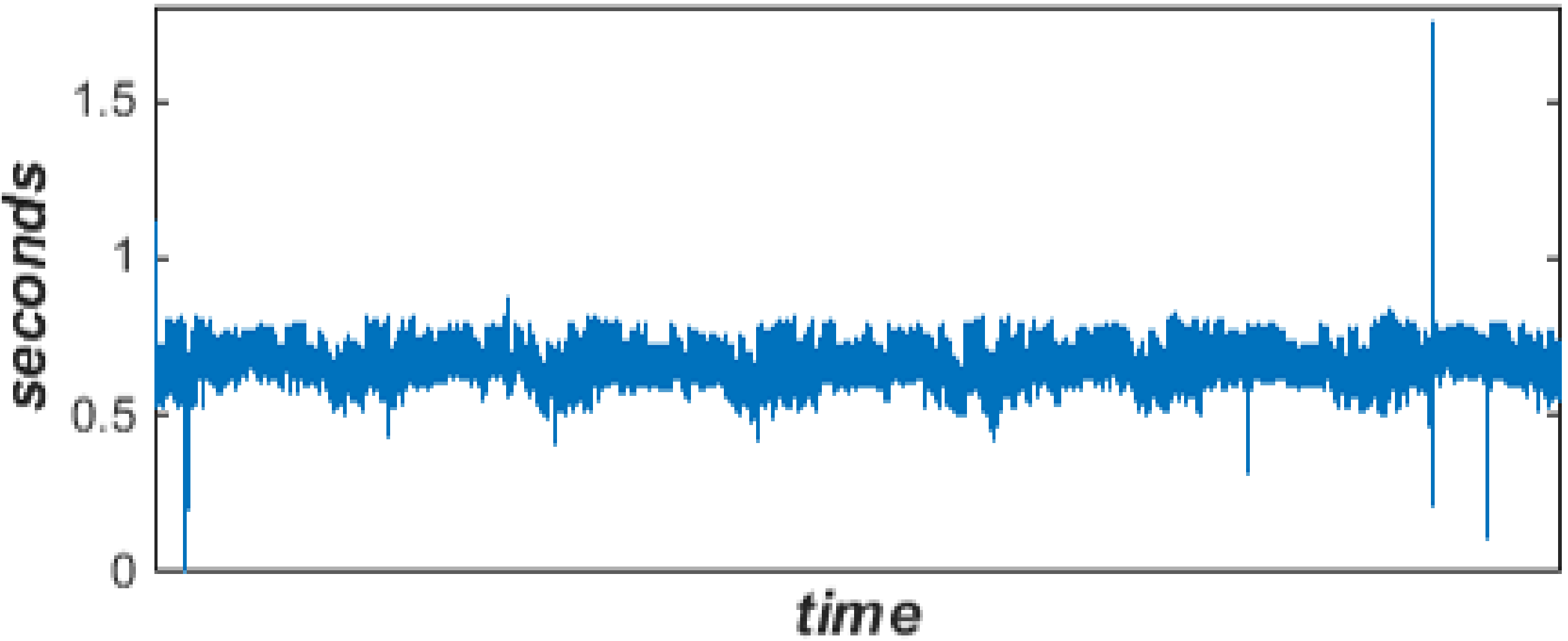}
\caption{Response time for on-demand auto-scaling}
\label{fig:reponse_time_on_demand}
\end{figure}

\begin{figure}
\begin{center}
\subfigure[$0\%$ on-demand resources]
{\label{fig:response_time_one_spot_zero}
\includegraphics[width=4in]{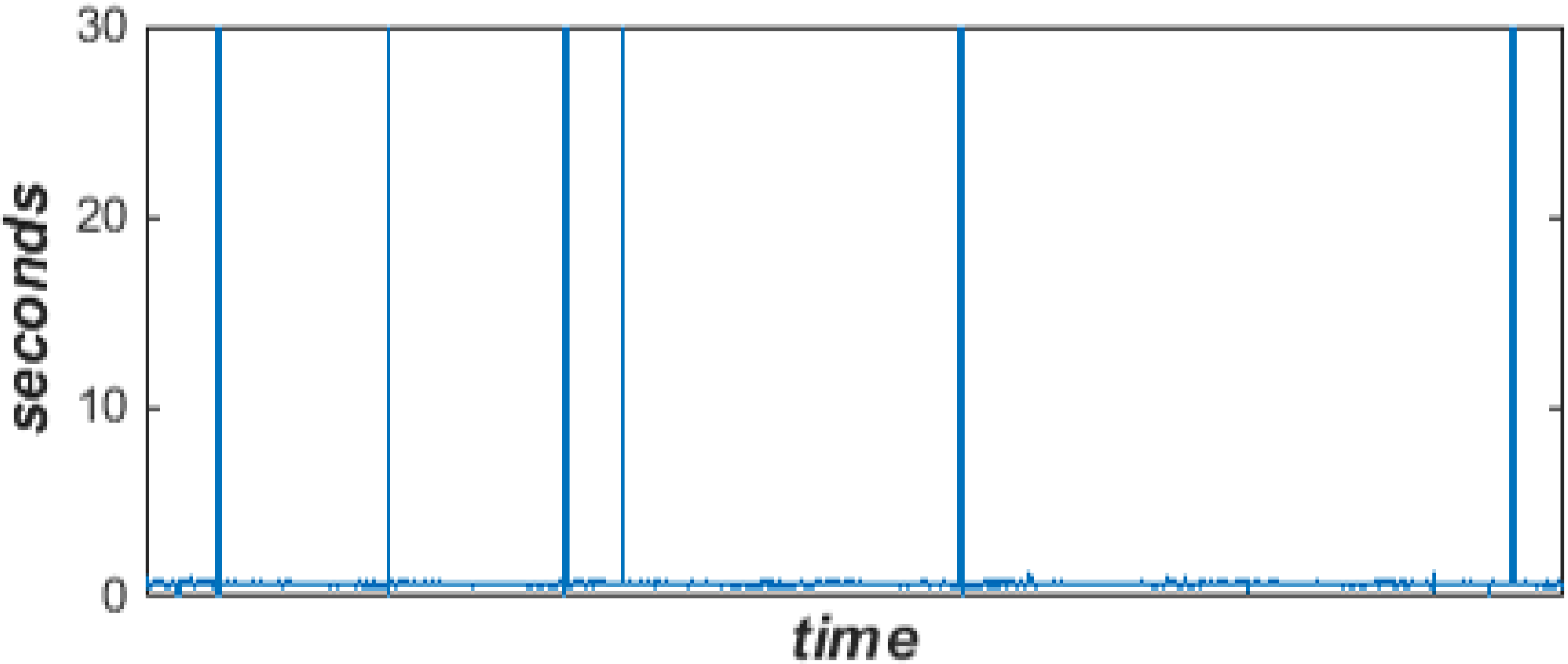}}
\subfigure[$20\%$ on-demand resources]
{\label{fig:response_time_one_spot_twenty}
\includegraphics[width=4in]{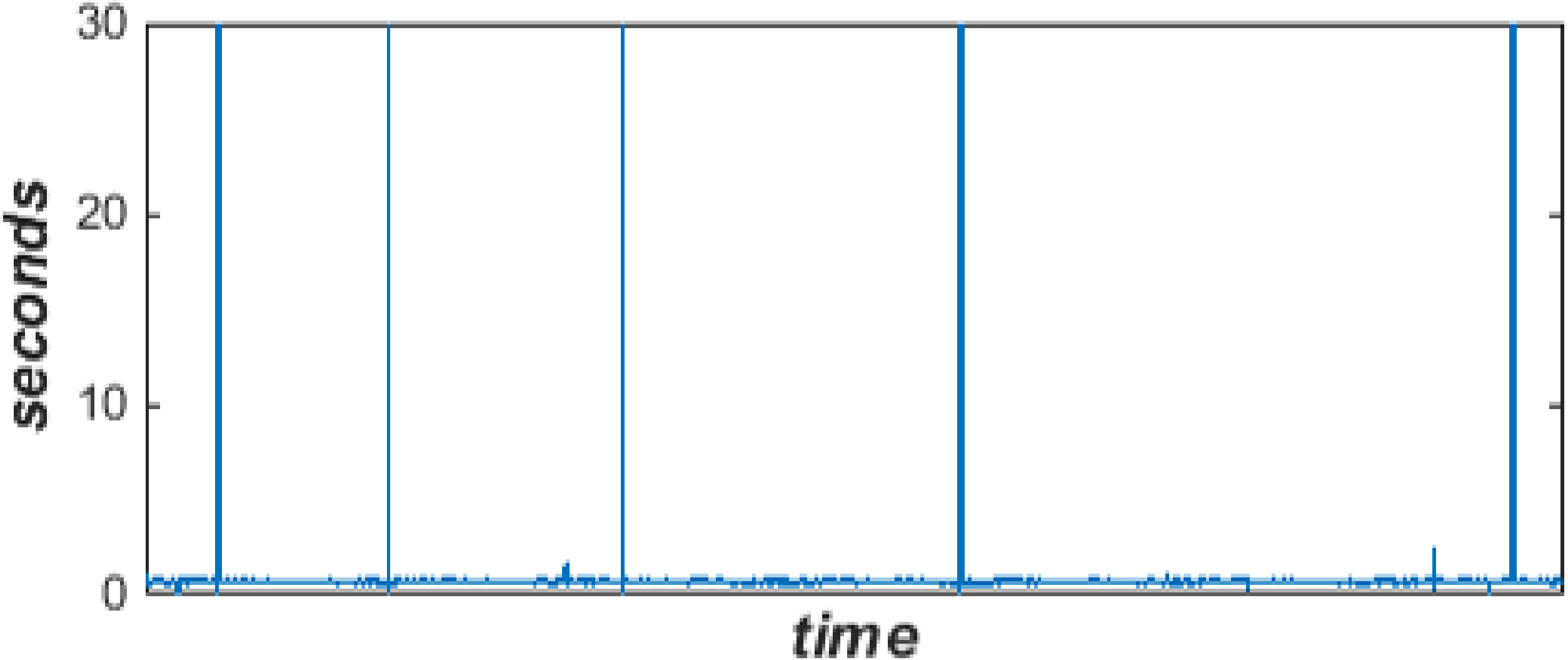}}
\subfigure[$40\%$ on-demand resources]
{\label{fig:response_time_one_spot_forty}
\includegraphics[width=4in]{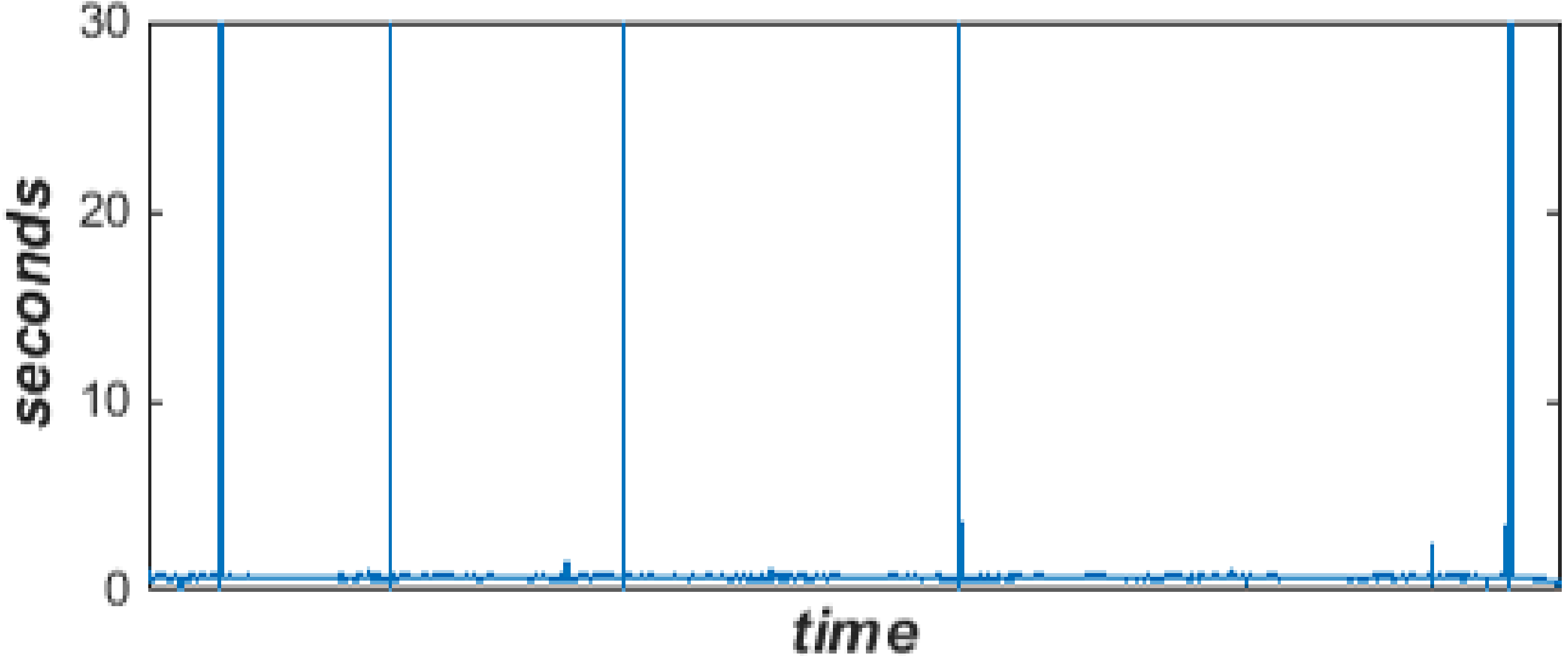}}
\end{center}
\caption{Response time of one spot type auto-scaling with various percentage of on-demand resources and truthful bidding strategy}
\label{fig:response_time_one_spot}
\end{figure}

\begin{figure}
\begin{center}
\subfigure[$0\%$ on-demand resources]
{\label{fig:response_time_ft_0_zero}
\includegraphics[width=4in]{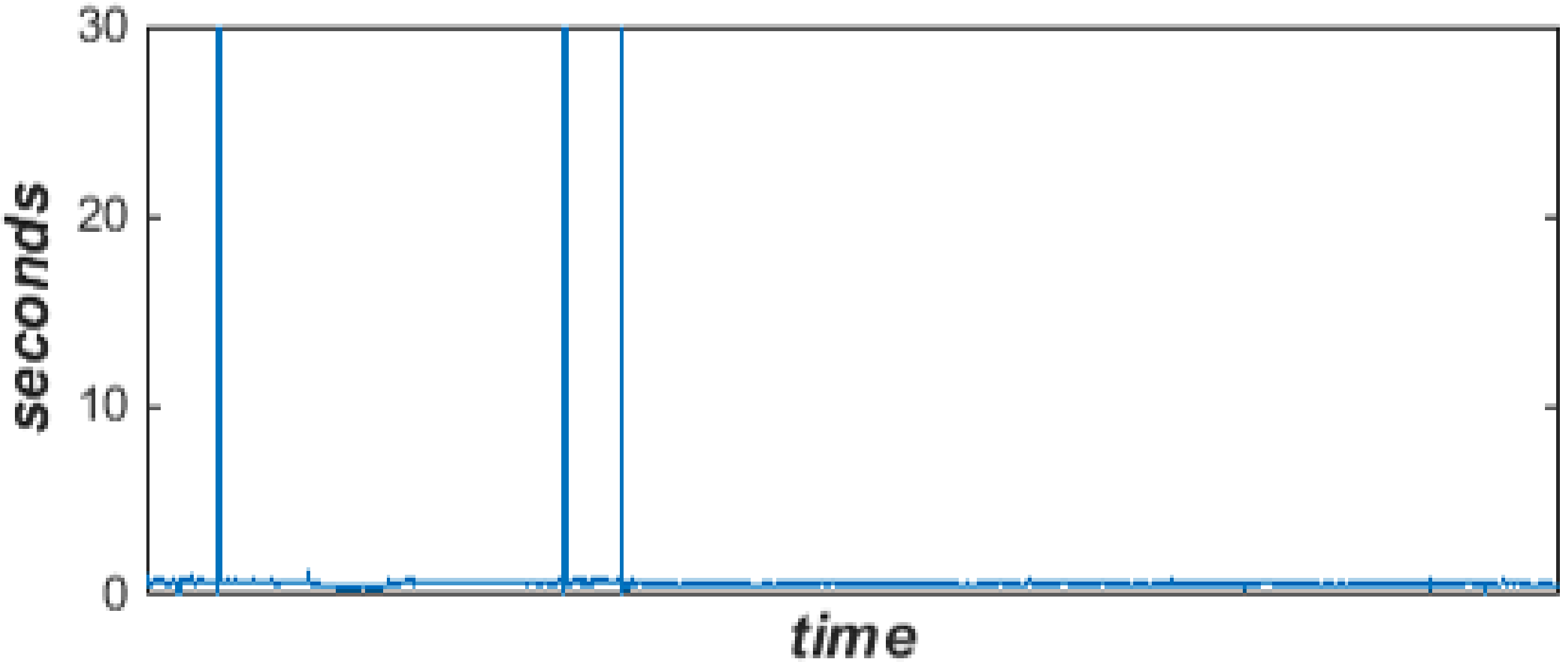}}
\subfigure[$20\%$ on-demand resources]
{\label{fig:response_time_ft_0_twenty}
\includegraphics[width=4in]{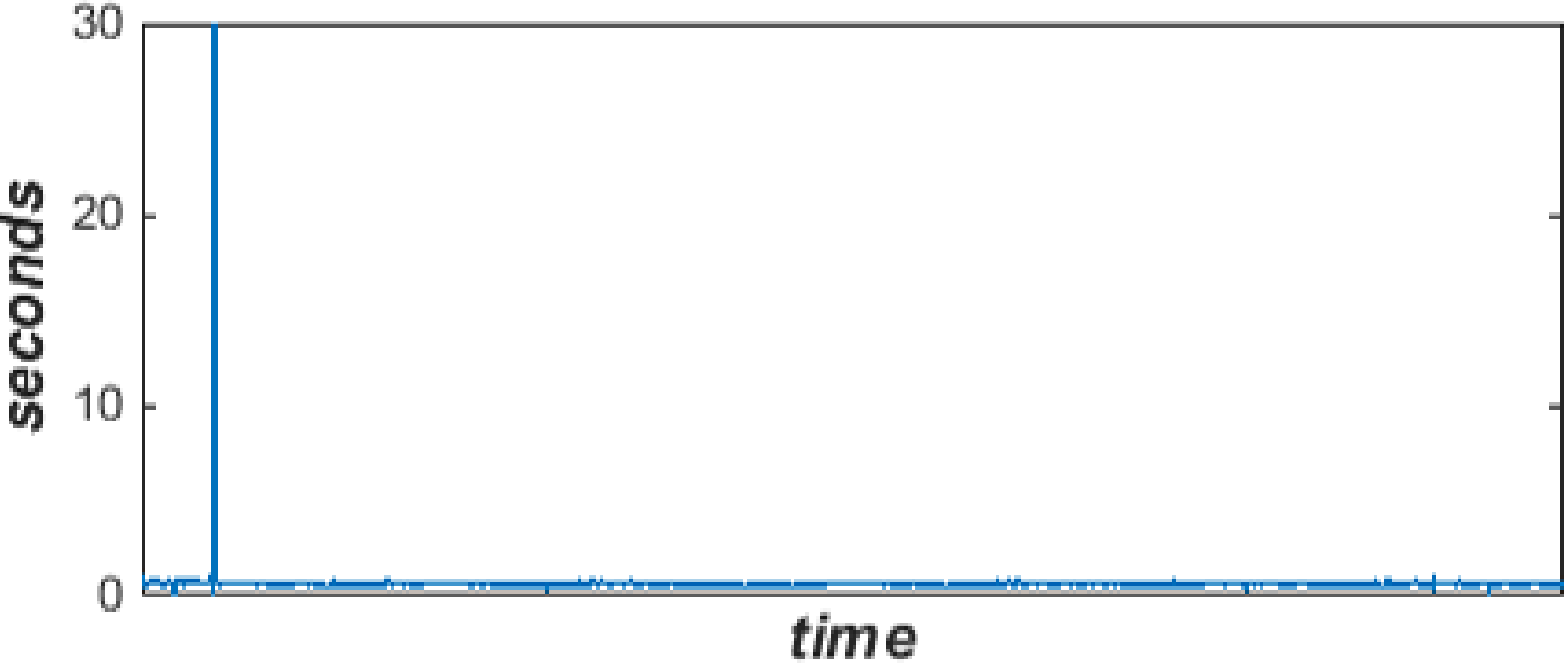}}
\subfigure[$40\%$ on-demand resources]
{\label{fig:response_time_ft_0_forty}
\includegraphics[width=4in]{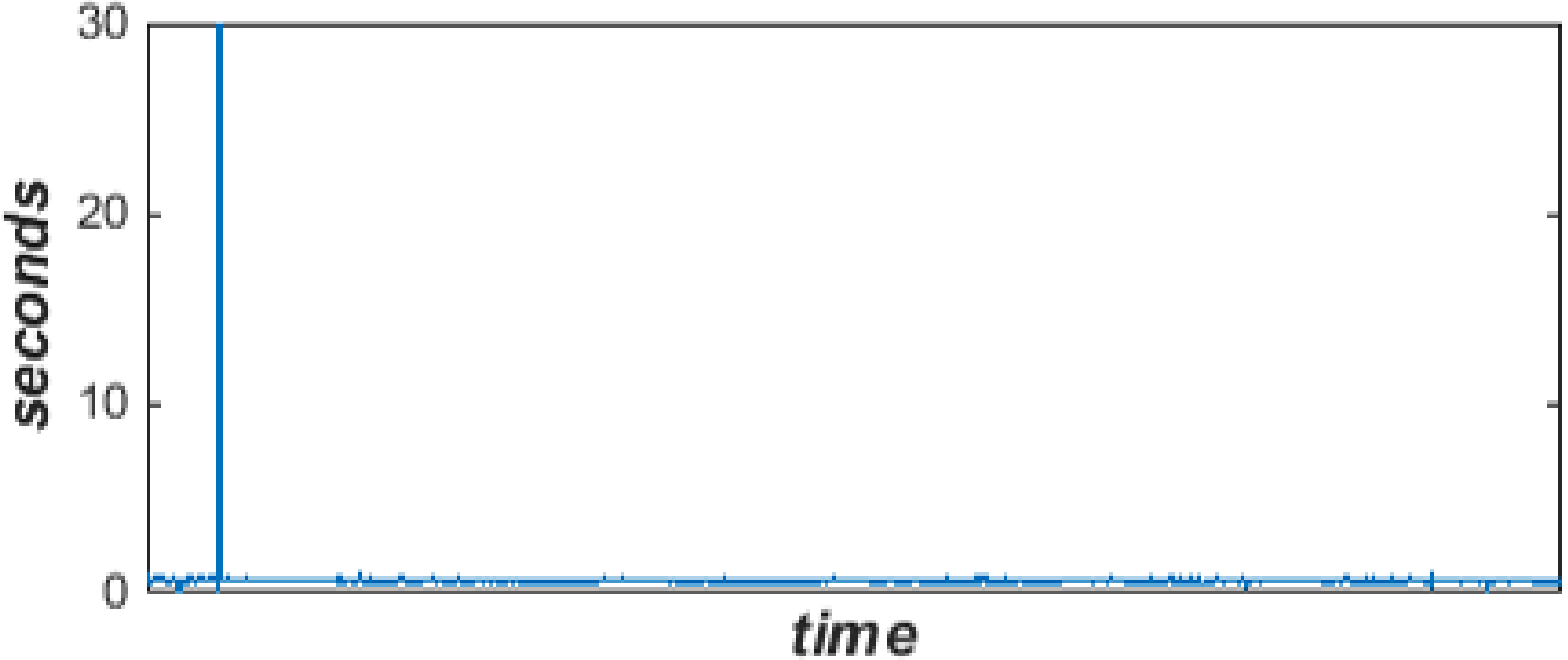}}
\end{center}
\caption{Response time of $f-0$ with various percentage of on-demand resources, truthful bidding strategy, and dynamic resource margin}
\label{fig:response_time_ft_0}
\end{figure}

\begin{figure}
\begin{center}
\subfigure[$0\%$ on-demand resources]
{\label{fig:response_time_ft_1_zero}
\includegraphics[width=4in]{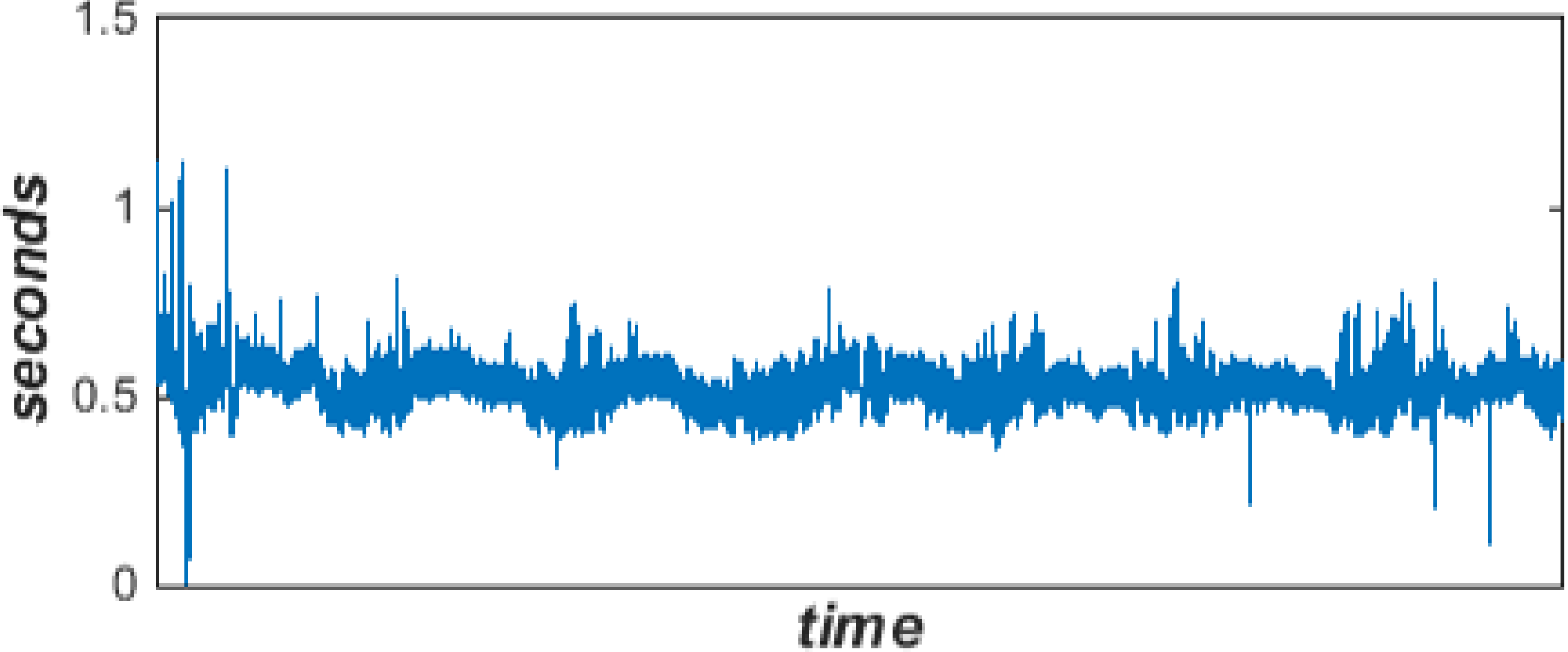}}
\subfigure[$20\%$ on-demand resources]
{\label{fig:response_time_ft_1_twenty}
\includegraphics[width=4in]{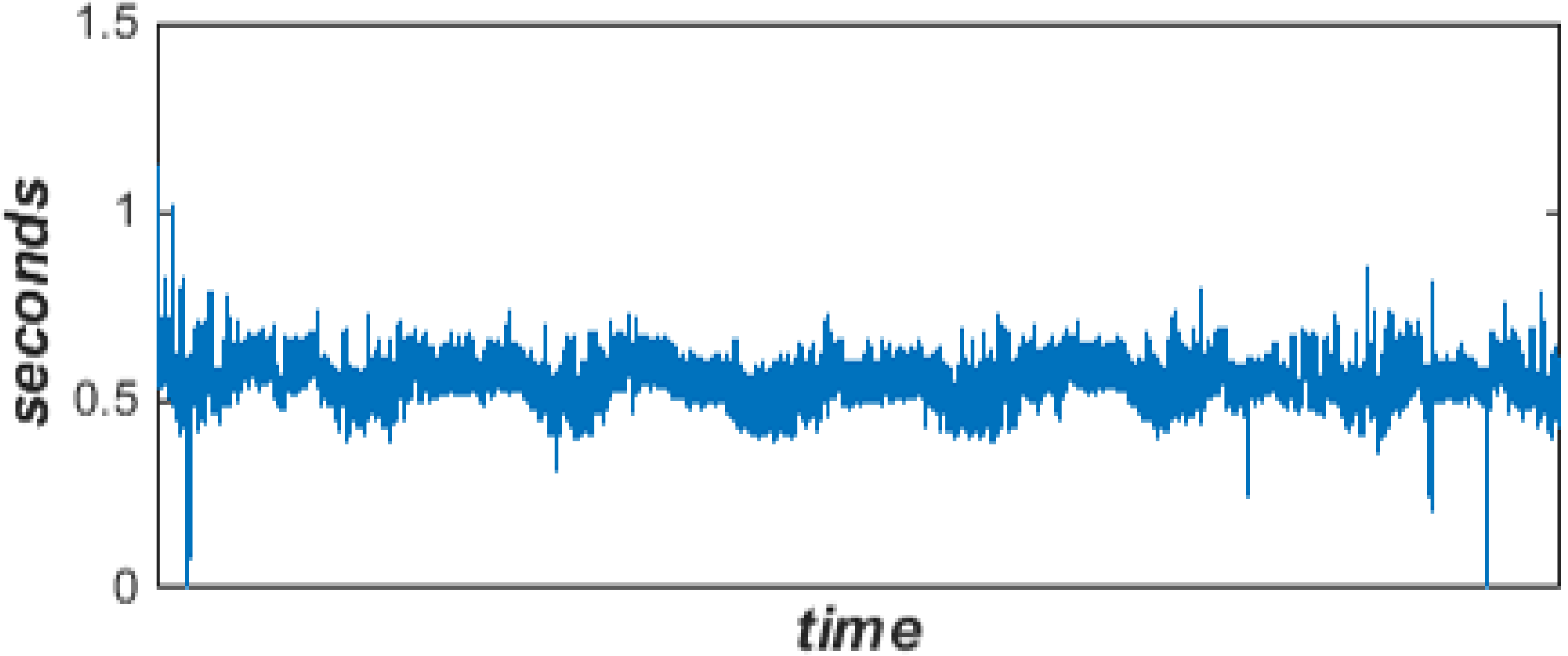}}
\subfigure[$40\%$ on-demand resources]
{\label{fig:response_time_ft_1_forty}
\includegraphics[width=4in]{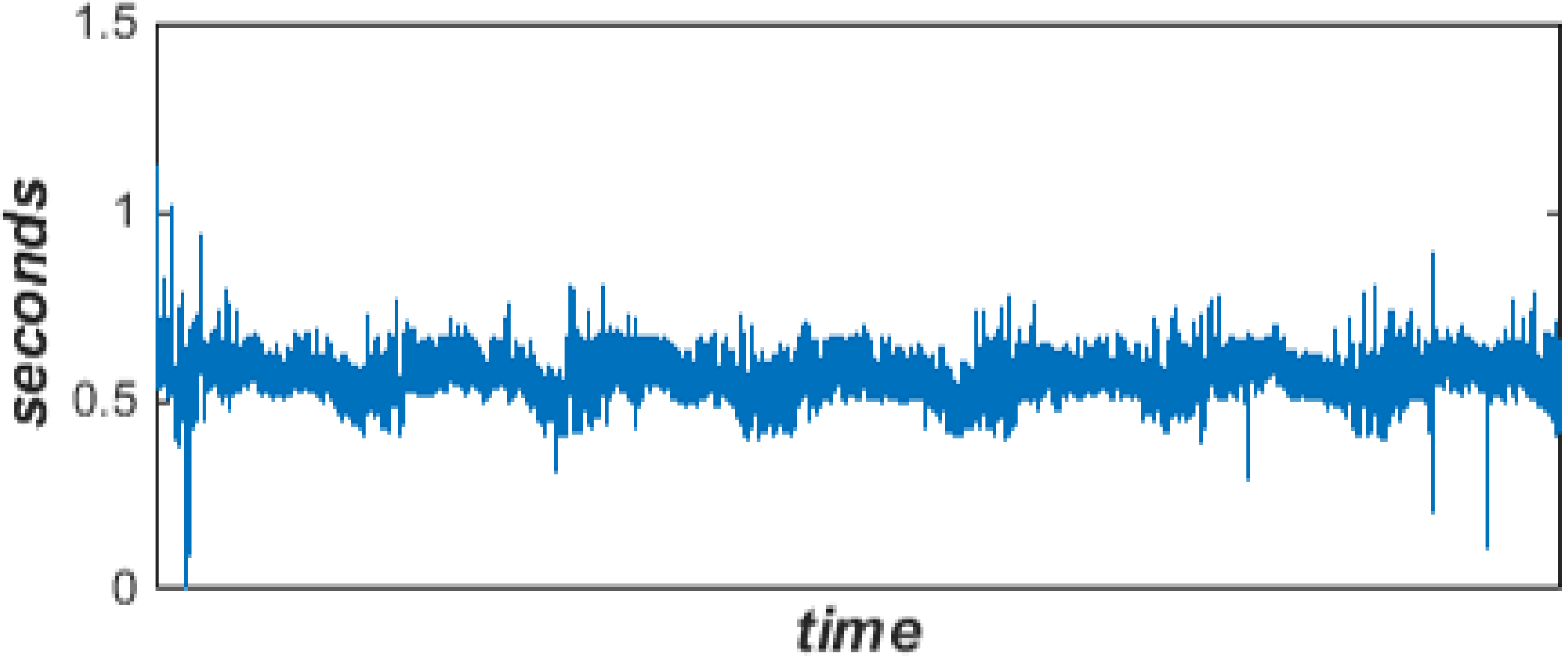}}
\end{center}
\caption{Response time of $f-1$ with various percentage of on-demand resources, truthful bidding strategy, and dynamic resource margin}
\label{fig:response_time_ft_1}
\end{figure}

\begin{figure}
\begin{center}
\subfigure[$0\%$ on-demand resources]
{\label{fig:response_time_one_spot_zero_on_demand_bidding}
\includegraphics[width=4in]{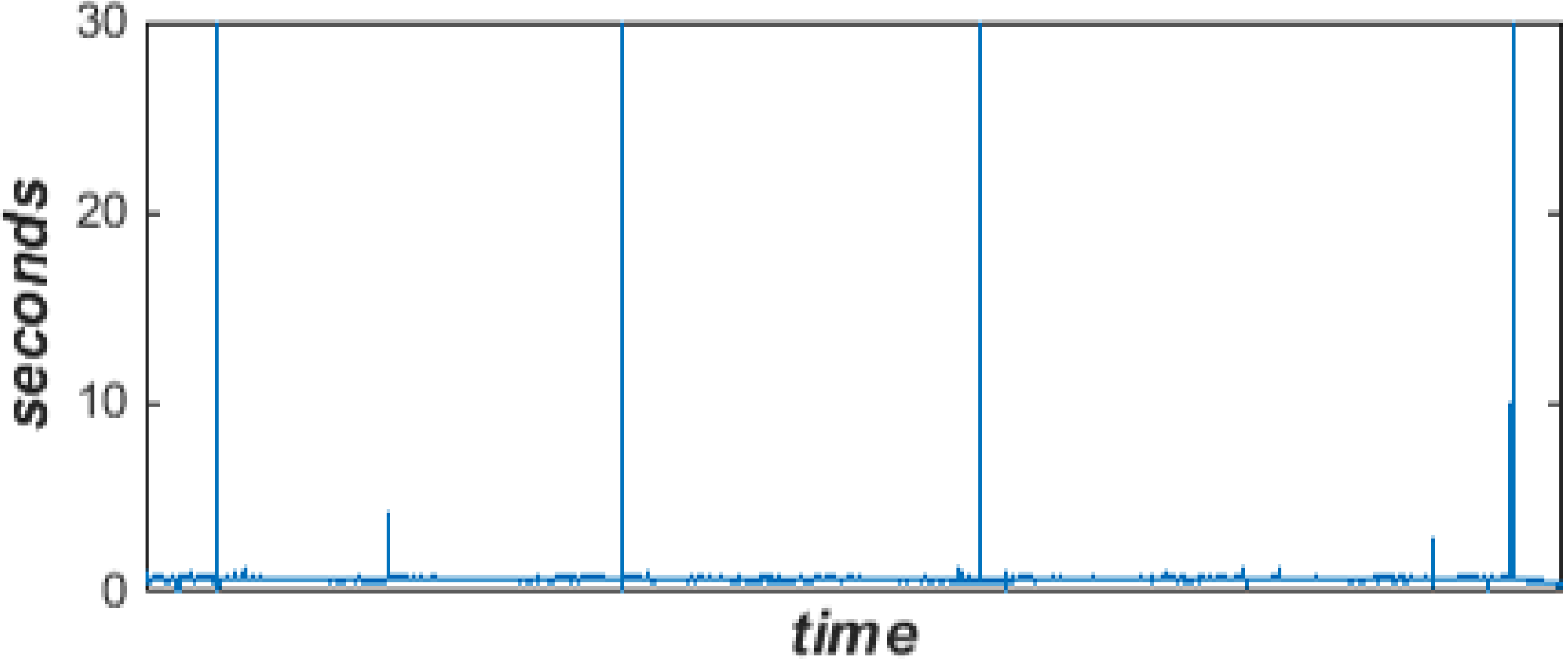}}
\subfigure[$20\%$ on-demand resources]
{\label{fig:response_time_one_spot_twenty_on_demand_bidding}
\includegraphics[width=4in]{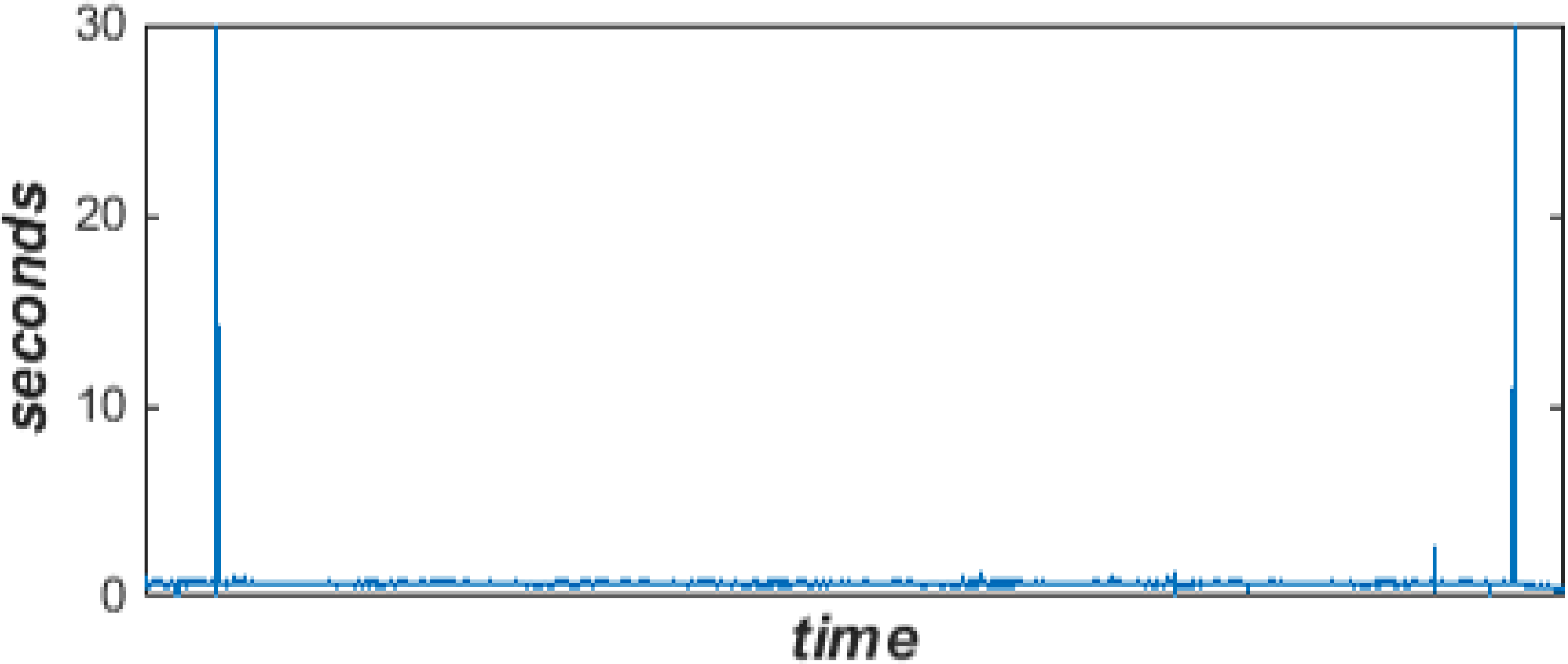}}
\subfigure[$40\%$ on-demand resources]
{\label{fig:response_time_one_spot_forty_on_demand_bidding}
\includegraphics[width=4in]{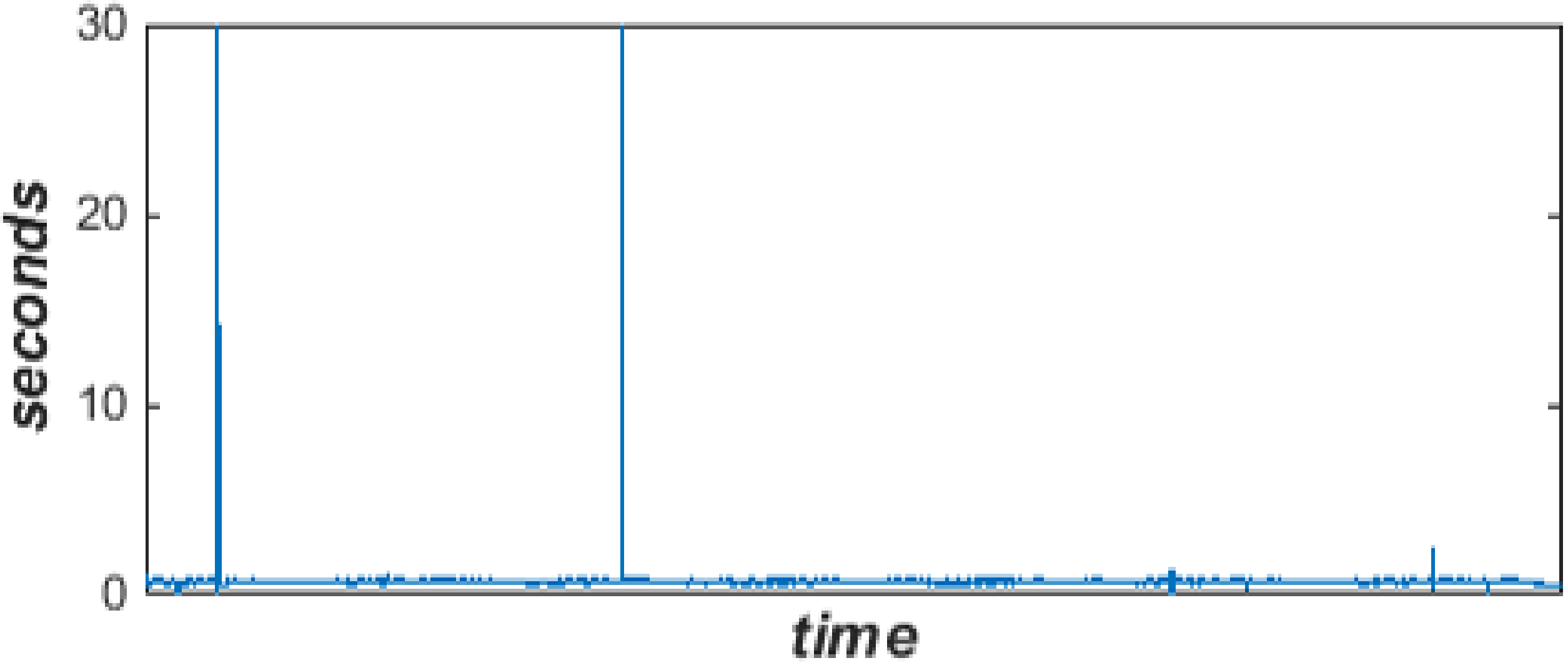}}
\end{center}
\caption{Response time of one spot type auto-scaling with various percentage of on-demand resources and on-demand bidding strategy}
\label{fig:response_time_one_spot_on_demand_bidding}
\end{figure}

\subsubsection{Response time}
Figure \ref{fig:reponse_time_on_demand}, \ref{fig:response_time_one_spot}, \ref{fig:response_time_ft_0}, and \ref{fig:response_time_ft_1} respectively depict real-time average response time of requests using on-demand, one spot, and our approach with truthful bidding strategy and dynamic resource margin. From the results, the on-demand auto-scaling produced smooth response time all along the experimental duration except for a peak that was caused by the corresponding peak in the workload. All experiments employing one spot type auto-scaling experienced periods of request timeouts caused by termination of spot instances, and only increasing the amount of on-demand resources cloud not improve the situation. While our approach greatly reduced such unavailability of service even using $f-0$ with no over-provision of resources. By using $f-1$, we were able to completely eliminate the timeouts under the recorded spot market traces. We omit the results for tests using $f-2$ and $f-3$ as they reveal similar results as Figure \ref{fig:response_time_ft_1}.

To show the effect of different bidding strategies, we compare the response time results of one spot type auto-scaling using the two proposed bidding strategies as they reveal the most significant difference. As Figure \ref{fig:response_time_one_spot} and Figure \ref{fig:response_time_one_spot_on_demand_bidding} present, it is obvious that service availability can be much improved with higher bidding prices using one spot type auto-scaling. On the other hand, the remaining timeouts also indicate that increasing bidding prices alone is not enough to guarantee high availability.

\subsubsection{Cost}

\begin{sidewaystable}
\scriptsize
\caption{Total Costs for Experiments with Various Configurations}
\label{tab:cost}
\begin{center}
\begin{tabular} {c | c | c | c | c}
\hline
\hline
\backslashbox{\textbf{Policies}}{\textbf{USD\$}} & \multicolumn{4}{c}{\textbf{Total Cost}} \\
\hline
\textbf{on-demand} & \multicolumn{4}{c}{116.34}\\
\hline
\hline
& \multicolumn{2}{|c|}{\textbf{Truthful Bidding}} & \multicolumn{2}{c}{\textbf{On-Demand Bidding}} \\
\hline
\textbf{one spot with $0\%$ on-demand} & \multicolumn{2}{|c|}{22.26} & \multicolumn{2}{c}{23.14} \\
\hline
\textbf{one spot with $20\%$ on-demand} & \multicolumn{2}{|c|}{46.50} & \multicolumn{2}{c}{47.33} \\
\hline
\textbf{one spot with $40\%$ on-demand} & \multicolumn{2}{|c|}{63.17} & \multicolumn{2}{c}{63.43} \\
\hline
\hline
\textbf{$f-0$ with $0\%$ on-demand} & \multicolumn{2}{|c|}{32.00} & \multicolumn{2}{c}{32.30} \\
\hline
\textbf{$f-0$ with $20\%$ on-demand} & \multicolumn{2}{|c|}{54.45} & \multicolumn{2}{c}{56.10} \\
\hline
\textbf{$f-0$ with $40\%$ on-demand} & \multicolumn{2}{|c|}{68.34} & \multicolumn{2}{c}{69.64} \\
\hline
\hline
 & \textbf{Static Resource Margin} & \textbf{Dynamic Resource Margin} & \textbf{Static Resource Margin} & \textbf{Dynamic Resource Margin}\\
\hline
\textbf{$f-1$ with $0\%$ on-demand} & 41.57 & 39.32 & 43.17 & 41.66 \\
\hline
\textbf{$f-1$ with $20\%$ on-demand} & 60.21 & 59.52 & 61.82 & 61.06 \\
\hline
\textbf{$f-1$ with $40\%$ on-demand} & 72.09 & 72.55 & 72.96 & 73.08 \\
\hline
\hline
\textbf{$f-2$ with $0\%$ on-demand} & 50.48 & 47.38 & 51.67 & 49.38 \\
\hline
\textbf{$f-2$ with $20\%$ on-demand} & 67.72 & 65.52 & 68.71 & 66.09 \\
\hline
\textbf{$f-2$ with $40\%$ on-demand} & 78.01 & 76.74 & 78.3 & 76.74 \\
\hline
\hline
\textbf{$f-3$ with $0\%$ on-demand} & 67.87 & 62.61 & 68.79 & 61.50 \\
\hline
\textbf{$f-3$ with $20\%$ on-demand} & 83.27 & 78.33 & 81.18 & 76.19 \\
\hline
\textbf{$f-3$ with $40\%$ on-demand} & 89.86 & 85.57 & 88.09 & 84.46 \\
\hline
\hline
\end{tabular}
\end{center}
\end{sidewaystable}

Table \ref{tab:cost} lists the total costs produced by all the experiments. Comparing to the cost of on-demand auto-scaling, we managed to gain significant cost saving using all other configurations. Tests using one spot type auto-scaling with $0\%$ on-demand resources realized the most cost saving up to $80.87\%$ regardless of its availability issue. 

The results show the amount of on-demand resources has a significant influence on cost saving. It also can be noted that higher fault-tolerant level incurs extra cost. Though optimal configuration of fault-tolerant level is always application specific, according to our results, configuration using $f-1$ with $0\%$ on-demand resource is the best choice under current market situation in regards of both financial cost and service availability. 

The resulted cost differences caused by different bidding strategies are generally small. Therefore, it is better to bid higher to improve availability if user's bidding has negligible impact on the market price.

As dynamic resource margin is only applicable when application is over-provisioned, we give the results for tests using dynamic resource margin when fault-tolerant level is higher than zero. According to the results, dynamic resource margin can bring extra cost saving and the amount of cost saving increases when more over-provision is necessary (i.e., higher fault-tolerant level). Though the resulted cost saving is not significant, it is safely achieved without sacrificing availability and performance of the application.

\subsection{Real Experiments}

We conducted two real tests on Amazon EC2 respectively using on-demand auto-scaling policies and the proposed auto-scaling policies with configuration of $f-1$ and $0\%$ on-demand. Other parameters are defined the same to the simulation tests.

We set up the experimental environment to run the Wikibench \cite{Baaren2009} benchmark tool. The major advantage of this tool compared to other tools such as TPC-W, RUBiS, and CloudStone is that it is stateless, which is the characteristic of modern highly scalable cloud services \cite{wilderchapter2012}. The tool is composed of three components:

\begin{itemize}
\item a client driver that mimics clients by continuously sending requests to the application server according to the workload trace;
\item a stateless application server installed with the Mediawiki application;
\item a mysql database loaded with the English Wikipedia data by the date of Jan 3rd, 2008.
\end{itemize}

\begin{figure}[!t]
\centering
\includegraphics[width=4in]{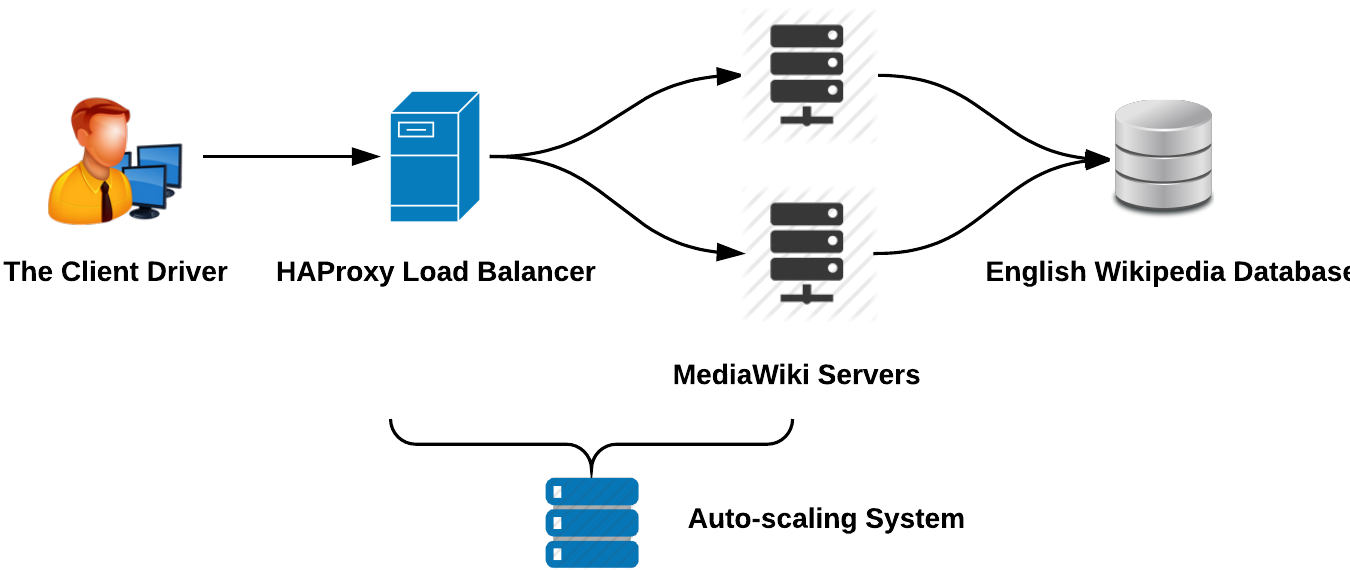}
\caption{The Testbed Architecture}
\label{fig:testbed}
\end{figure}

\noindent Our aim is to scale the application-tier. Thus, we inserted a HAProxy load balancer layer into the original architecture in order to let the client driver talk to a cluster of servers. The architecture of the testbed is illustrated in Figure \ref{fig:testbed}. We picked the first 3 days of the Wikipedia workload \cite{Baaren2015,Urdaneta2009} (Figure \ref{fig:workload}) and scaled it down to half of its original rate as the workload for testing because Amazon limits the number of instances each account can launch.

The testing environment resided in Amazon \emph{\textbf{us-east-1d}} zone which is in a relatively busy region with higher degree and frequency of price fluctuations. Regarding each component, we launched one \emph{c4.large} instance acting as the client driver, one \emph{m3.medium} instance running the HAProxy load balancer, and one \emph{c4.2xlarge}\footnote{The 4th generation instances were introduced between the time we performed the simulations and the real experiments. To be consistent, we only consider the 13 spot types listed in Figure \ref{fig:spot_price_history} for both the simulations and the real experiments} instance serving the mysql database requests. The auto-scaling system itself is running on a local desktop computer remotely in Melbourne. Before the tests, we profiled each component to make sure none of them become the bottleneck of the system.

The test using the proposed approach started at 3:30am September 9, 2015, Wednesday, US east time. Its testing period spanned across three busy weekdays from Wednesday to Friday.

\begin{figure}[!t]
\centering
\includegraphics[width=4in]{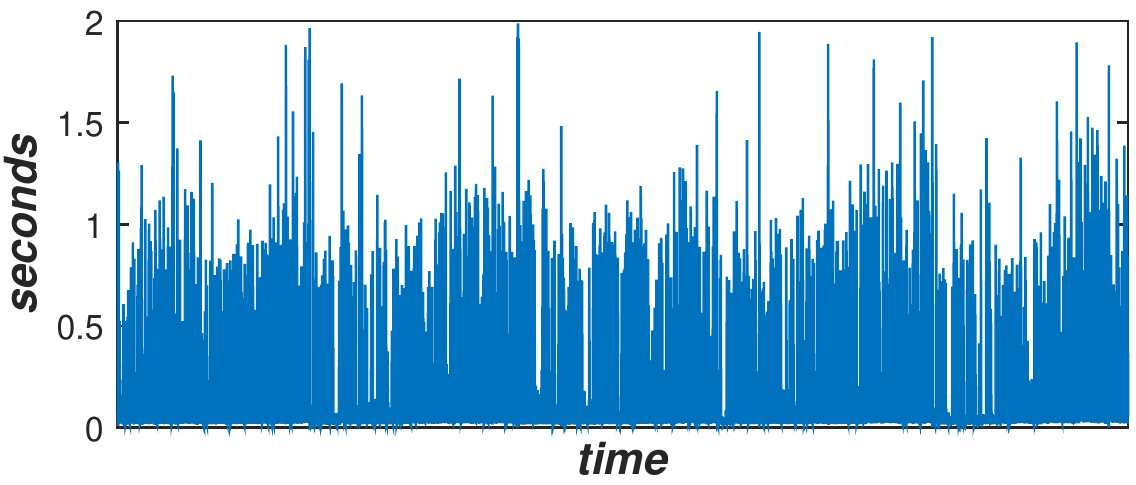}
\caption{Response time for on-demand auto-scaling on Amazon}
\label{fig:reponse_time_on_demand_amazon}
\end{figure}

\begin{figure}[!t]
\centering
\includegraphics[width=4in]{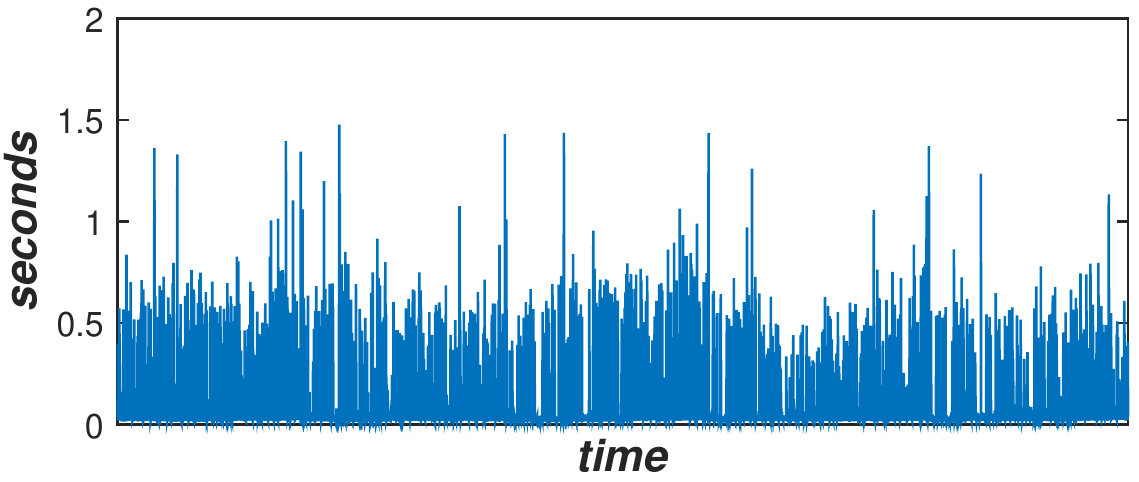}
\caption{Response time for spot auto-scaling on Amazon}
\label{fig:response_time_spot_amazon}
\end{figure}

Figure \ref{fig:reponse_time_on_demand_amazon} and \ref{fig:response_time_spot_amazon} presents real-time response time results of the two experiments. Both results suffer from peaks of high response time. By studying the recorded log, we confirmed they were not caused by shortage of resources as resource utilizations of all the involving VMs were never beyond safe threshold during both tests. Various other reasons can be the culprits, such as cold cache, short term network issues, interference from the shared virtualized environment, and garbage collection \cite{Dean2013}. 
We encountered three unexpected terminations during the test of our approach. Thanks to the fault-tolerant mechanism and policies, we managed to avoid service interruption and performance degradation during those periods. 
In addition, because resources are tighter in on-demand auto-scaling, it generally performs worse in response time compared to the proposed approach.

\begin{table}
\caption{Cost of the Experiments}
\label{tab:real_cost}
\begin{center}
\begin{tabular} {c | c}
\hline
\hline
 & \textbf{Cost(USD\$)}\\
\hline
\textbf{on-demand} & 19.01\\
\hline
\textbf{$ft-1$ and $0\%$ on-demand} & 5.69\\
\hline
\end{tabular}
\end{center}
\end{table}

Regarding cost, we calculated the total cost of application servers in both experiments. Table \ref{tab:real_cost} presents the results. The proposed approach reaches $70.07\%$ cost saving.

\subsection{Discussion}

Even with high fault-tolerant level, the proposed approach cannot guarantee $100\%$ availability, and no solution can ever manage to assure absolute service continuity due to the nature of spot market. What our system offers is a best effort to counter large scale surges of market prices of the selected spot types in a short time, which is highly unlikely under current market condition. In fact, we have not encountered any case that more than one spot group fail simultaneously during simulations, real experiments, and testing phases. However, market condition could change. Hence, application provider should adjust configuration of the auto-scaling system dynamically according to real-time volatility of the spot market. In addition, the nature of the application also affects the decision. If the application is availability-critical, higher fault-tolerant level is always desirable. Adversely, for some applications, such as analytical jobs, even one spot type auto-scaling is acceptable. 

The presented results in Section \ref{sec:performance_evalutaion} only indicates the cost saving potential of a certain application considering a selected set of spot types under the recorded spot market prices and workload traces. Thanks to the dynamic truthful bidding price mechanism, even in competitive market condition, we can ensure that the cost reduction gained by our approach will not vanish but only diminish. To reach more cost saving, the application provider can take into account a broader set of spot types, which is available in Amazon's offering.

To save cost and time for testing, application providers can tune the parameters of the auto-scaling system in a similar way as we did by first utilizing simulation for fast validation and then test the system in production environment.

There are also differences in price among the same spot types across different availability zones. It is trivial to extend the current fault-tolerant model to utilize spot groups from multiple availability zones. Currently, the auto-scaling system limits the selection of spot groups within the same availability zone due to charges for traffic across availability zones. If the application provider has already adopted a multi-availability-zone deployment, such extension is able to realize more cost saving.

The overhead of the auto-scaling system is negligible. As presented in Section \ref{sec:scaling_policies}, the time complexity of the scaling policies is not significant. The frequency that the scaling policies are called depends on the monitoring interval and the frequency of price changes, which are at least in the scale of seconds.
\section{Related Work}
\label{sec:related_work}

\subsection{Horizontally Auto-scaling Web Applications}
Horizontally auto-scaling web applications have been extensively studied and applied \cite{Lorido-Botran2014}. Basically, auto-scaling techniques for web applications can be classified into three categories: \emph{reactive approaches}, \emph{proactive approaches}, and \emph{mixed approaches}. Reactive approaches scale applications in accordance of workload changes. Proactive approaches predicts future workload and scale applications in advance. Mixed approaches can scale applications both reactively and proactively.

Most industry auto-scaling systems are reactive-based. Among them, the most frequently used service is Amazon's Auto Scaling Service \cite{Amazon2015}. It requires user to first create an auto-scaling group, which specifies the type of VMs and image to use when launching new instances. Then user should define his scaling policies as rules like ``add 2 instances when CPU utilization is larger than 75\%''. Another popular service is offered by RightScale. Their service is based on a voting mechanism that lets each running instance decide whether it is necessary to grow or shrink the size of the cluster based on their own condition \cite{RightScale2015}.

Other than just using simple rules to make scaling decisions, researchers have developed scaling systems based on formal models. These models aim to answer the question that how many resources are actually required to serve certain amount of incoming workload under QoS constraints. Such model can be simply obtained using profiling techniques as we did in this paper. Other commonly adopted approaches include queueing models  \cite{Urgaonkar2008,Jiang,Han2014,Gandhi2014,Gandhi,Salah2015} that either abstract the application as a set of parallel queues or a network of queues, and online learning approaches such as reinformacement learning \cite{Dutreilh,Barrett2013,Xiangping2013}.

Proactive auto-scaling is desirable because time taken to start and configure newly started VMs creates a resource gap when workload suddenly surges to the level beyond capability of the available resources. To satisfy strict SLA, sometimes it is necessary to provision enough resources before workload actually rises. As workloads of web applications usually reveal temporal patterns, accurate prediction of future workload is feasible using state-of-art time-series analysis and pattern recognition techniques. A lot of them have been applied to auto-scaling of web applications \cite{Jiang,Roy,Jingqi,Caron2011,Islam2012,Wei,Herbst2014,Dutta}.

Most auto-scaling systems only utilize homogeneous resources. While some, including our system, have explored using heterogeneous resources to provision web applications. Upendra et al. \cite{Upendra}, and Srirama and Ostavar \cite{Srirama} adopt integer linear programming (ILP) to model the optimal heterogeneous resource configuration problem under SLA constraints. Fernandez et al. \cite{Fernandez} utilizes tree paths to represent different combinations of heterogeneous resources and then searches the tree to find the most suitable scaling plan according to user's SLA requirements.

Different from the above works, our objective goes beyond using minimum resources to provision the application. Instead, we want to devise fault-tolerant mechanism and auto-scaling policies that comply with the fault-tolerant semantics to reliably scale web applications on cheap spot instances. We believe the reviewed auto-scaling techniques are complementary to our approach. The proposed system can incorporate their resource estimation models, and workload prediction techniques as well.

\subsection{Application of Spot Instances}
There have been a lot of attempts to use spot instances to cut resource cost under various application context. Resource provision problems using spot instances have been studied for fault-tolerant applications \cite{Costache,Binnig,Poola2014,Sifei,Voorsluys,Changbing,Chohan,Zafer,Hsuan-Yi,Subramanya2015} such as high performance computing, data analytics, MapReduce, and scentific workflow.

For these applications, the fault-tolerant mechanism is often built on checkpointing, replication, and migration. Multiple novel checkpointing mechanisms \cite{Jangjaimon2015,Sangho2012,Jung2011} have been developed to allow these applications to harness the power of spot instances. SpotOn \cite{Subramanya2015} combines multiple fault-tolerant mechanisms to increase the cost-efficiency and performance of batch processing applications running on spot instances.

Regarding web applications, Han et al. \cite{Han} proposed a stochastic algorithm to plan future resource usage with a mixture of on-demand and spot instances. Except they only use homogeneous resources, their problem is also different to ours as they aim to plan the resource usage with the knowledge of the future while we provision resources dynamically. Mazzucco and Dumas \cite{Mazzucco} also explored using a mixture of homogeneous on-demand instances and spot instances to provision web applications. Instead of building a reliable auto-scaling system, their target is to maximize web application provider's profit by using an admission control mechanism at the front end to dynamically adapt to sudden changes of available resources.

Sharma at al. proposed a derivative IaaS cloud platform based on spot instances called SpotCheck \cite{Singh2014,Sharma2015}. To transparently provide high availability on spot instances to end users, they incorporated technologies, such as nested virtualization, live VM migration, and time-bounded VM migration with memory checkpointing, to dynamically move users' VMs when underlying spot instances are available or revoked. Because of its transparency to end users, it is ideal for cloud brokers and large organizations with high resource demands. While our approach is lightweight and thus more suitable for small organizations who want to harness the power of spot instances by themselves. He et al. \cite{He2015} from the same group evaluated the ability of the approach to reliably run web applications on spot instances. Though they do not provision redundant capacity as we do, they reported non-negligible overhead incurred by nested virtulization. Their proposed system \cite{Singh2014,Sharma2015,He2015} is able to preserve the memory state of the revoked spot VMs, which enables it to seamlessly host stateful applications. Though our approach requires the application to be stateless, this does not reduce its generality as highly scalable cloud applications are expected to be stateless \cite{wilderchapter2012}, and stateful applications can be easily turned into stateless by storing session information in a memory cache cluster \cite{wilderchapter2012}. 
Their system relies on the termination warnings issued by existing providers \cite{Amazonb} to be able to conduct migrations in time. Our approach is capable of operating in possible future spot markets that do not provide termination warnings.

Recently, Amazon EC2 introduced a new feature, called Spot Fleet API \cite{Amazona}. It allows user to bid for a fixed amount of capacity possibly constituted by instances of different spot types. It continuously and automatically provisions the capacity using the combination of instances that incurs the lowest cost. However, as its provision decision ignores reliability, it is not suitable to provision web applications. 

\section{Conclusions and Future Work}
\label{sec:conclusion}
In this paper, we explored how to reliably and cost-efficiently auto-scale web applications using a mixture of on-demand and heterogeneous spot instances. We first proposed a fault-tolerant mechanism that can handle unexpected spot terminations using heterogeneous spot instances and over-provision. We then devised novel cost-efficient auto-scaling policies that comply with the defined fault-tolerant semantics for hourly-billed cloud market. We implemented a prototype of the proposed auto-scaling system on Amazon EC2 and a simulation version on CloudSim \cite{Calheiros2011} for repeatable and fast validation. We conducted both simulations and real experiments to demonstrate the efficacy of our approach by comparing the results with the benchmark approaches.

In the future, we plan to further optimize our system by incorporating the following features:

\begin{itemize}
\item selection of spot groups according to predicted spot prices in near future;
\item dynamic decision of fault-tolerant level and proportion of on-demand instances according to volatility of the spot market using machine leaning technologies;
\item an interface that allows web application providers to plug in different workload prediction techniques into the auto-scaling system to achieve proactive auto-scaling; and
\item utilization of spot groups across different availability zones.
\end{itemize}

\section*{Acknowledgment}

We thank the anonymous reviewers, Adel Nadjaran Toosi, Xunyun Liu, Yaser Mansouri, Bowen Zhou, and Nikolay Grozev for their valuable comments in improving the quality of the paper. This work is supported by Australian Research Council Future Fellowship.

\bibliographystyle{elsarticle-num}
\bibliography{spot_auto_scaling_copy}

\end{document}